\documentclass[fleqn,usenatbib]{mnras}

\usepackage{newtxtext,newtxmath}
\usepackage[T1]{fontenc}
\usepackage{ae,aecompl}

\usepackage{graphicx}	% Including figure files
\usepackage{amsmath}	% Advanced maths commands
\usepackage{scalerel}
\usepackage{tikz}
\usepackage{mathtools}

\usetikzlibrary{svg.path}

\definecolor{orcidlogocol}{HTML}{A6CE39}
\tikzset{
  orcidlogo/.pic={
    \fill[orcidlogocol] svg{M256,128c0,70.7-57.3,128-128,128C57.3,256,0,198.7,0,128C0,57.3,57.3,0,128,0C198.7,0,256,57.3,256,128z};
    \fill[white] svg{M86.3,186.2H70.9V79.1h15.4v48.4V186.2z}
                 svg{M108.9,79.1h41.6c39.6,0,57,28.3,57,53.6c0,27.5-21.5,53.6-56.8,53.6h-41.8V79.1z M124.3,172.4h24.5c34.9,0,42.9-26.5,42.9-39.7c0-21.5-13.7-39.7-43.7-39.7h-23.7V172.4z}
                 svg{M88.7,56.8c0,5.5-4.5,10.1-10.1,10.1c-5.6,0-10.1-4.6-10.1-10.1c0-5.6,4.5-10.1,10.1-10.1C84.2,46.7,88.7,51.3,88.7,56.8z};
  }
}

\newcommand\orcidicon[1]{\href{https://orcid.org/#1}{\mbox{\scalerel*{
\begin{tikzpicture}[yscale=-1,transform shape]
\pic{orcidlogo};
\end{tikzpicture}
}{|}}}}
\newcommand{\CO}{$^{12}$CO}		                                    %writes 12CO
\newcommand{\COthird}{$^{13}$CO}	                             	%writes 13CO
\newcommand{\molh}{H$_2$}		                                  	%writes H2
\newcommand{\xco}{$X_{\mathrm{CO}}$}	                        	%writes xco
\newcommand{\avxco}{$\left\langle X_{\mathrm{CO}}\right\rangle$}    %writes <xco>

\title[A convergence study of synthetic CO emission]{Synthetic CO emission and the $X_{\rm CO}$ factor of young molecular clouds: a convergence study}

\author[Borchert et al.]{E.~M.~A. Borchert\orcidicon{0000-0002-6994-8874},$^{1,2}$\thanks{E-mail: elisabeth.borchert@monash.edu}
	S.~Walch\orcidicon{0000-0001-6941-7638},$^{2,3}$
	D.~Seifried\orcidicon{0000-0002-0368-9160},$^{2}$
	S.~D.~Clarke\orcidicon{0000-0001-9751-4603},$^{2}$
	A.~Franeck,$^{2,4}$
    \newauthor P.~C.~N\"{u}rnberger\orcidicon{0000-0003-4384-3115}$^{2}$
	\\
	$^{1}$School of Physics and Astronomy, Monash University, Clayton Vic 3800, Australia\\
	$^{2}$I. Physikalisches Institut, Universit\"at zu K\"oln, Z\"ulpicher Str. 77, 50937 K\"oln, Germany
	\\
	$^{3}$Center for Data and Simulation Science, Universit\"at zu K\"oln, www.cds.uni-koeln.de, Germany
	\\
	$^{4}$Astronomical Institute, Czech Academy of Sciences, Bocni II 1401, CZ-141 00 Prague, Czech Republic
}

\date{Accepted XXX. Received YYY; in original form ZZZ}

\pubyear{2021}

\begin{document}
	\label{firstpage}
	\pagerange{\pageref{firstpage}--\pageref{lastpage}}
	\maketitle
	
	% Abstract of the paper
	\begin{abstract}
		The properties of synthetic CO emission from 3D simulations of forming molecular clouds are studied within the SILCC-Zoom project. Since the time scales of cloud evolution and molecule formation are comparable, the simulations include a live chemical network. Two sets of simulations with an increasing spatial resolution (\mbox{d$x=3.9$ pc} to \mbox{d$x=0.06$ pc}) are used to investigate the convergence of the synthetic CO emission, which is computed by post-processing the simulation data with the \textsc{radmc-3d} radiative transfer code. To determine the excitation conditions, it is necessary to include atomic hydrogen and helium alongside {\molh}, which increases the resulting CO emission by $\sim7-26$ per cent. Combining the brightness temperature of $^{12}$CO and {\COthird}, we compare different methods to estimate the excitation temperature, the optical depth of the CO line and hence, the CO column density. An intensity-weighted average excitation temperature results in the most accurate estimate of the total CO mass. When the pixel-based excitation temperature is used to calculate the CO mass, it is over-/underestimated at low/high CO column densities where the assumption that $^{12}$CO is optically thick while {\COthird} is optically thin is not valid. Further, in order to obtain a converged total CO luminosity and hence {\avxco} factor, the 3D simulation must have \mbox{d$x\lesssim0.1$ pc}. The {\avxco} evolves over time and differs for the two clouds; yet pronounced differences with numerical resolution are found. Since high column density regions with a visual extinction larger than 3~mag are not resolved for d$x\gtrsim 1$~pc, in this case the {\molh} mass and CO luminosity both differ significantly from the higher resolution results and the local {\xco} is subject to strong noise.
		Our calculations suggest that synthetic CO emission maps are only converged for simulations with d$x\lesssim 0.1$ pc.
	
	\end{abstract}
	
	% Select between one and six entries from the list of approved keywords.
	% Don't make up new ones.
	
	%%%%%%%%%%%%%%%%%%%%%%%%%%%%%%%%%%%%%%%%%%%%%%%%%%
	
	%%%%%%%%%%%%%%%%% BODY OF PAPER %%%%%%%%%%%%%%%%%%
	
	%%%%%%%%%%%%%%%%%%%% INTRODUCTION %%%%%%%%%%%%%%%%%%
	
	\section{Introduction}
	
	Molecular clouds (MCs) are of great interest as they are regions where star formation takes place. As molecular hydrogen ({\molh}) is difficult to observe due to its lack of a permanent dipole moment as well as widely spaced rotational energy levels \citep{Glover11}, it is usually observed indirectly. Common approaches to do so are i) the use of carbon monoxide (CO) and the conversion to H$_2$ mass via the so-called $X_{\rm CO}$ factor \citep{Bolatto13}, ii) radiative transfer models to obtain the column density of a molecular species from which the abundance relative to {\molh} is estimated using chemical models (e.g. \citealt{Dickman78, Dickman86} for {\COthird} and \citealt{Frerking82, Bachiller86, Cernicharo87} for other CO isotopologues), iii) the use of dust extinction maps \citep[e.g.][]{Lombardi01, Lombardi06}, iv) the use of dust emission maps \citep[e.g. from Herschel observations][and many more]{Koenyves10, Arzoumanian11, Schneider12}.
	
	To derive the column density of e.g. CO directly, one has to consider that with increasing column density,  \mbox{{\CO}(J=$1\rightarrow0$)} quickly becomes optically thick. In fact, most of the observable CO emission originates from optically thick areas (as we show later in this paper). Hence, it is necessary to combine {\CO} observations with an optically thin tracer in order to determine the optical depth, excitation temperature, and then column density of CO. Often {\COthird} and ${\rm C^{18}O}$ are used for this purpose \citep[see][and many others]{Pineda08, Pineda10, Arzoumanian13, Kong18}.
	
	On galactic scales, the {\xco} factor is commonly used to determine the molecular gas content from {\CO} observations \citep[see e.g.][for a review]{Bolatto13}. It is a constant of proportionality, which converts the integrated \mbox{{\CO}(J=$1\rightarrow0$)} emission, $I_{\mathrm{CO}}$\footnote{For convenience, we drop the superscript in {\CO} from hereon and just write CO.}, to the {\molh} column density, $N_{\mathrm{H_2}}$, as \mbox{$N_{\mathrm{H_2}} = X_{\mathrm{CO}} \times I_{\mathrm{CO}}$}. A typical value for the Milky Way is \mbox{$ X_{\mathrm{CO,MW}}=2\times 10^{20}\:{\rm cm}^{-2}\:({\rm K\: km\: s}^{-1})^{-1}$} \citep{Bolatto13}.
	
	In a resolved region, an average {\xco} factor, {\avxco}, can be determined from
	\begin{equation}
	\left\langle X_{\mathrm{CO}}\right\rangle=\frac{\sum N_{\mathrm{H_2}}}{\sum I_{\mathrm{CO}}},
	\label{eq:Xco}
	\end{equation}
	summing over all pixels of the map. 
	However, a single {\xco} does not work well for resolved molecular clouds. 
	Previous numerical models (\citealt{Shetty11a, Shetty11b}; \citealt{Clark15}; \citealt{Glover16}; \citealt{Szuecs16}; \citealt{Seifried17b, Seifried20}; \citealt{Penaloza18}) and observational results \citep{Lee14, Lewis20} have shown that there are strong spatial variations of the factor of up to an order of magnitude. \cite{Gong20} investigate the variation of the {\xco} factor as a function of the galactic environment and find the {\xco} factor is decreasing with increasing metallicity and increasing cosmic ray ionisation rate while the factor stays unaffected by the far-ultraviolet radiation field (the so-called interstellar radiation field, ISRF). Furthermore they find that {\xco} is at first decreasing with increasing density, as the excitation temperature increases, and then see it increasing again when the CO emission is fully saturated. Another part that the factor misses out on is the so-called CO-dark gas, where {\molh} is present but no CO (\citealt{Lada88}; \citealt{Grenier05}; \citealt{Wolfire10}). \cite{Grenier05} concluded through gamma ray emission that $\gtrsim 30$ per cent of the {\molh} gas is CO-dark.
	
	Previous studies of {\xco} using synthetic observations of numerical simulations suggest that a numerical resolution of 2 pc is sufficient to derive a converged {\xco} factor \citep{Gong18}. This is in conflict with previous findings from \cite{Seifried17b} and \cite{Joshi19}, who show that the CO abundance, and consequently the synthetic CO emission, is not converged as long as the effective resolution is below $\sim$0.1 pc in their 3D simulations.
	
	In this paper we analyse the synthetic CO emission of forming molecular clouds from the SILCC-Zoom project modelled with different (increasing) spatial resolutions (\citealt{Walch15}; \citealt{Girichidis16}; \citealt{Seifried17b}). The synthetic emission maps are calculated with the radiative transfer code \textsc{radmc-3d} \citep{Dullemond2012}.
	
	Furthermore we investigate the accuracy of using the CO and {\COthird} intensities to estimate the CO column densities.
	
	This paper is structured in the following way. \mbox{Section \ref{sec:SILCC}} briefly describes the SILCC-Zoom simulations used in this paper. In \mbox{Section \ref{sec:RT}} we give a short overview of the radiative transfer method and its related parameters. In \mbox{Section \ref{sec:column_density}} we use the synthetic emission maps of CO and {\COthird} to estimate the CO column density using various assumptions for the excitation temperature and compare our findings to the column density present in the simulation. \mbox{Section \ref{sec:emission_analysis}} shows a resolution study of the synthetic emission from CO, leading to the findings of the {\xco} factor which we discuss and analyse in more detail in \mbox{Section \ref{sec:Xco}}. We close the paper with our conclusions in \mbox{Section \ref{sec:Conclusion}}.
	
	%%%%%%%%%%%%%%%%%%%% SILCC-SIMULATION %%%%%%%%%%%%%%%%%%
		\begin{table}
		\centering
		\label{tab:refinement}
		\begin{tabular}{c|c|c}
			\hline
			Name & Cloud & d$x$ [pc]\\ 
			\hline
			\hline 
			\textit{MC1\_L5} & MC2 & 3.9 \\ 
			\textit{MC1\_L6} & MC1 & 2.0 \\ 
			\textit{MC1\_L7} & MC1 & 1.0 \\ 
			\textit{MC1\_L8} & MC1 & 0.5 \\ 
			\textit{MC1\_L9} & MC1 & 0.24 \\ 
			\textit{MC1\_L10} & MC1 & 0.12 \\ 
			\textit{MC1\_L11} & MC1 & 0.06 \\ 
			& & \\
			\textit{MC2\_L5} & MC2 & 3.9 \\ 
			\textit{MC2\_L6} & MC2 & 2.0 \\ 
			\textit{MC2\_L7} & MC2 & 1.0 \\ 
			\textit{MC2\_L8} & MC2 & 0.5 \\ 
			\textit{MC2\_L9} & MC2 & 0.24 \\ 
			\textit{MC2\_L10} & MC2 & 0.12 \\ 
			\textit{MC2\_L11} & MC2 & 0.06 \\
			\hline
		\end{tabular} 
		\caption{List of the SILCC-Zoom simulations used for the radiative transfer post-processing. We give the run name (column 1), which consists of the molecular cloud complex (column 2) and the maximum refinement level (column 3). The latter corresponds to the effective resolution d$x$ given in column 3.}
	\end{table}
	
	\section{SILCC-Zoom simulations}\label{sec:SILCC}
	The data analysed in this work is based on the SILCC \citep{Walch15, Girichidis16} and SILCC-Zoom simulations \citep{Seifried17b}, which model the evolution of the multi-phase interstellar medium and the life-cycle of molecular clouds from spatial scales of 500 pc down to $\sim 0.1$ pc. The simulations are carried out with the adaptive mesh refinement (AMR) code \mbox{{\sc Flash} 4.3} \citep{Fryxell00}. Within {\sc Flash}, we solve the \mbox{(magneto-)hydrodynamical} equations with a 2$^{\rm nd}$ order finite-volume scheme (i.e. here we use hydrodynamical simulations carried out with the Bouchut 5-wave magneto-hydrodynamics solver with a zero magnetic field, ${\bf B} =0$; \citealt{Bouchut07, Waagan09, Bouchut10}). Further, we include the gravity of the stellar disc via an external potential, while the gas self-gravity is modelled with an OctTree-based method \citep{Wunsch18}. In addition, we modify {\sc Flash} to include a non-equilibrium chemical network following seven species: $\mathrm{CO, C^+, H_2, H, H^+, O\:and\:e^-}$ \citep[][]{NL97, Glover07b, Glover10, Walch11, Walch15}; and a method to compute the (self-)shielding of the forming molecular gas with the {\sc OpticalDepth} module \citep[presented in][]{Wunsch18} similar to the TreeCol method \citep{Clark12}. 
	
	The chemical network also evaluates the heating and cooling processes and calculates the gas and dust temperatures \citep[see][]{Glover10, Glover12}. The cooling processes we consider include the line cooling from $\text{C}^+$ and oxygen fine structure lines, ro-vibrational lines of $\text{H}_2$ and $\text{OH}$, and Lyman-$\alpha$. Also the energy transfer from gas to dust cools the gas. For temperatures ${T>10^{4}\,\mathrm{K}}$, helium and metals are assumed to be in collisional ionisation equilibrium and their cooling rates are tabulated in \citet{gnat2012ion}. %
	
	The initial condition is a uniform gas surface density of \mbox{$\Sigma_{\rm gas} = 10\;{\rm M}_{\odot}\;{\rm pc}^{-2}$}, a stellar surface density (used to determine the external gravitational potential) of \mbox{$\Sigma_{\star} = 30\;{\rm M}_{\odot}\;{\rm pc}^{-2}$}, and the uniform ISRF is set to 1.7 times the Habing field ($G_0 = 1.7$). These conditions are typical for the solar neighbourhood.
	
	The simulation is advanced for $t_0=11.9$ Myr up to which supernova (SN) explosions drive turbulence and a multi-phase ISM is established. Half of the 15 SN/Myr are randomly positioned in the $x-y-$plane, while the other half are placed in density peaks \citep[i.e. mixed SN driving;][]{Walch15}. Before the first molecular clouds are about to form, we start to zoom in at $t=t_0$ and continue the SILCC-Zoom simulations for another 4 Myr. This implies that we allow the code to adaptively refine in two pre-defined zoom-in volumes. Turbulent driving by SNe is switched off at this time to investigate the cloud evolution without further influence from external stellar feedback. Each zoom-in volume contains a molecular cloud complex called {\it MC1} and {\it MC2}, respectively. The simulation time $t$ used in the following corresponds to the difference between the total simulation time, $t_{\rm S}$, and $t_0$. More details on the SILCC-Zoom simulations can be found in \citet{Seifried17b}.

	\begin{figure*}
		\includegraphics[width=\textwidth]{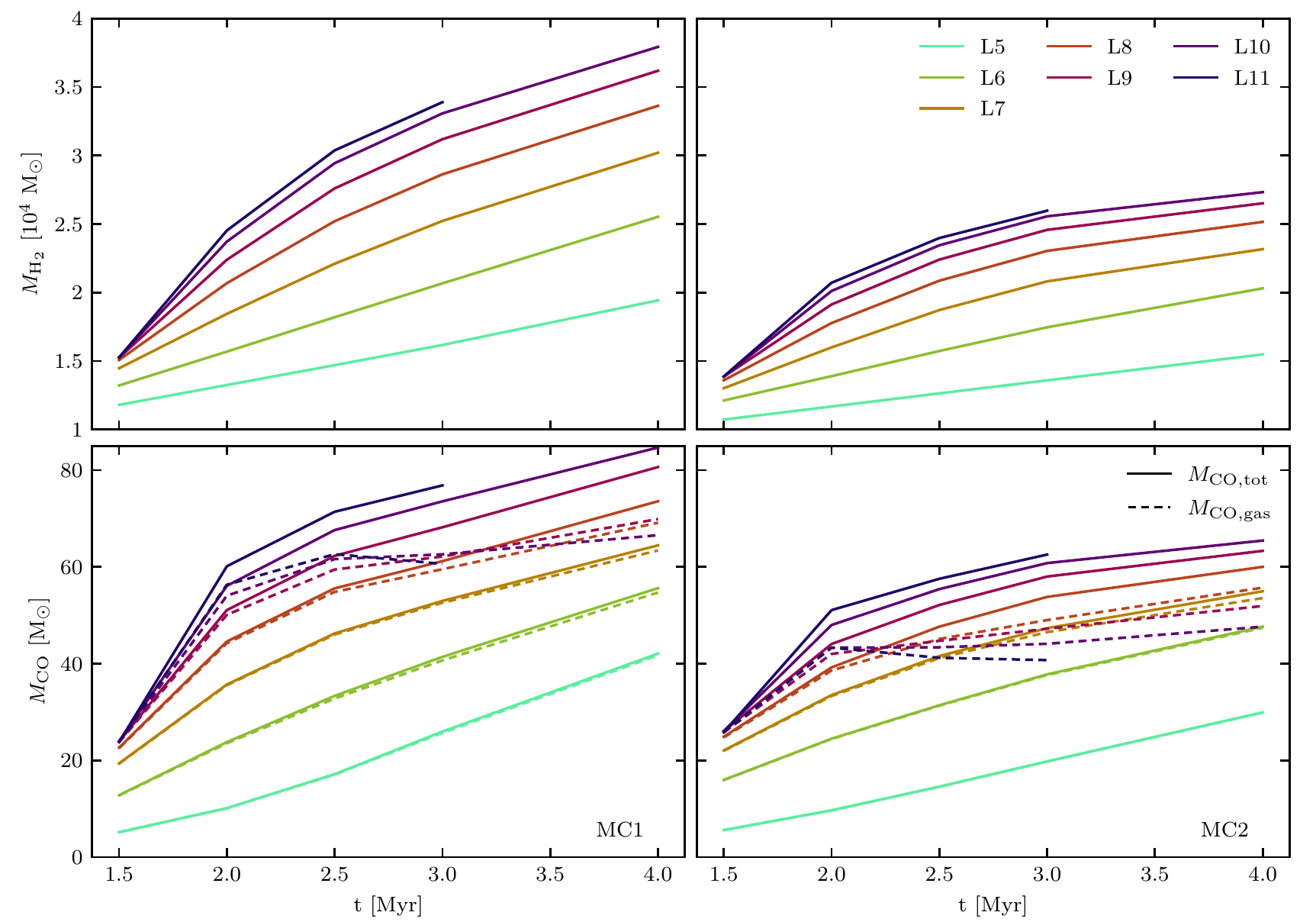}
		\caption{Time evolution ($t= t_{\rm S}-t_0$) of the total {\molh} mass (top) and CO mass (bottom) for the different maximum refinement levels. {\it Lx=L5} corresponds to 3.9 pc while {\it Lx=L11} corresponds to 0.06 pc (see also Table \ref{tab:refinement}). The left column shows MC1 while the right one shows MC2. The solid lines show the total mass while the dashed lines in the bottom panels show the remaining gas-phase CO mass after the simulation data has been post-processed with a CO freeze-out model.}
		\label{fig:M_co_h2}
	\end{figure*}

	The base grid of the SILCC simulation has a uniform linear cell size of 3.9 pc (corresponding to {\it Lx=L5}). Each increasing refinement level decreases the linear cell size by a factor of 2, such that {\it Lx=L11} corresponds to an effective resolution of 0.06 pc. Starting from the same initial condition, we carry out seven simulations for each molecular cloud complex, each one reaching a higher maximum refinement level (i.e. {\it Lx=L5...L11}). Otherwise the simulations are exactly the same. We list all simulations in Table \ref{tab:refinement}.
	
	We discuss the implications of the maximum refinement level on the formation of molecular gas (H$_2$ and CO) in \citet{Seifried17b}, where we show that an effective resolution of $\lesssim 0.1$ pc is necessary to obtain converged CO mass fractions (see also \citealt{Joshi19}). In Fig.~\ref{fig:M_co_h2}, we depict the time evolution\footnote{As the highest refinement level of the L11 run is not yet reached at 1.5 Myr but at 1.65 Myr, and due to its computational cost, we only analyse the three time steps, 2.0, 2.5 and 3.0 Myr in this case.} of the molecular gas mass (top panels: H$_2$ mass, $M_{\rm H_2}$; bottom panels: CO mass, $M_{\rm CO}$) for the different resolution runs as calculated from our on-the-fly chemical network. A molecular gas mass of a few $\times 10^4\;{\rm M}_{\odot}$ is typical for molecular clouds in the solar neighbourhood \citep[see e.g.][ from the Hershel Gould Belt Survey]{Bontemps10, Polychroni13, Arzoumanian19}. In this paper we show the effect of the resolution on the synthetic observations of CO and the derived $X_{\rm CO}$ factor. We will show that {\xco} is not yet converged at a resolution of d$x$=1~pc.

	CO is only emitting cooling radiation while in its gas-phase. Yet, CO quickly freezes out onto dust grains and is depleted from the gas phase. Thus, we post-process the simulation data with a CO freeze-out model based on \cite{Glover16} before computing the synthetic emission maps. We have previously applied the freeze-out model to calculate the gas-phase abundance of CO in isolated, dense molecular cloud filaments \citep{Seifried17a}. The remaining gas-phase CO mass is significantly smaller than the total CO mass, as illustrated by the dotted lines in Fig.~\ref{fig:M_co_h2}. The difference between gas-phase and total $M_{\rm CO}$ mass increases with increasing {\it Lx} as the colder and denser regions become better resolved. For {\it L10} this starts out at $\sim 8-10$ per cent at 2 Myr and increases to $\sim23-27$ per cent at 4 Myr.
	
	%%%%%%%%%%%%%%%%%%%% RADIATIVE TRANSFER POST-PROCESSING %%%%%%%%%%%%%%%%%%
	
	\section{Radiative transfer post-processing}\label{sec:RT}
	For the post-processing of the simulations we use the radiative transfer code \textsc{radmc-3d}, version 0.41 \citep{Dullemond2012}, a fully self-consistent radiative transfer code. The code can perform dust continuum and line radiative transfer simulations in 1D, 2D and 3D. As we are using \textsc{radmc-3d} to calculate the emission of CO and {\COthird}, we briefly outline what we require to perform our radiative transfer for the molecular lines and the mode we have used. 

	In order to work out the line transfer on an adaptive mesh (as present in our simulations), the AMR structure is given to \textsc{radmc-3d} in an Oct-tree format. We have recently written a python-based pipeline which can extract the Oct-tree from any {\sc Flash} simulation. Here we apply the so-called {\it FLASH-PP pipeline}\footnote{\url{https://bitbucket.org/pierrenbg/flash-pp-pipeline/src/master/}} to the SILCC-Zoom simulation data (see Section~\ref{sec:SILCC}). Our pipeline delivers the \textsc{radmc-3d} input Oct-tree and the input gas velocity, gas temperature, total density, as well as the number densities of CO and {\COthird} and each collisional partner (see Section~\ref{sec:collpartner}) in the required format. The number densities of H, {\molh}, and CO are directly extracted from the simulation data (as calculated by the chemical network). Here we apply the freeze-out model to get the number density of CO which is in the gas-phase, based on \cite{Glover16}. Using the constant isotopic ratio $\mathrm{CO}/\mathrm{^{13}CO}=69$ \citep{Wilson99} the {\COthird} number density is obtained. We have previously investigated the impact of a variable isotopic ratio on our {\COthird} column density, as suggested by \citet{Szuecs14}, and have found very little variations within the area classified as observable (see below for the classification). As pointed out by \cite{Liszt98} and \cite{Liszt20}, a decreased abundance ratio of CO/{\COthird} is expected for \mbox{$N(CO)\lesssim 5\times 10^{15}\:{\rm {cm}^{-2}}$} and $I_{\rm CO}\lesssim 5\:{\rm K\: km\: s^{-1}}$. Within our observable area the CO column density is typically larger than $10^{16}\:{\rm {cm}^{-2}}$.	Hence we do not further discuss it here.
	
	In \textsc{radmc-3d}, in order to calculate the CO emission, the level population has to be first obtained in every 3D grid cell. This can in principal be done by assuming local thermal equilibrium (LTE), where the kinetic temperature is used to calculate the level population in CO. In a dense medium, this is a good approach. However, in dilute gas this assumption breaks down as the excitation temperature differs from the gas temperature. In these regions, the excitation temperature is determined by the collisions of CO with other atoms and molecules in the gas, such as H$_2$, H, He (see Sec.~\ref{sec:collpartner} below). We aim to calculate the CO emission for a wide range of gas densities, and therefore apply the latter approach.
	
	Our post-processing method is done in non-LTE using the Large Velocity Gradient (LVG) \citep{Sobolev57,Ossenkopf97} and an Escape Probability method\footnote{This is lines\_mode=3 in \textsc{radmc-3d}}. LVG was first used in \textsc{radmc-3d} by \cite{Shetty11a}, who include a description of the method used in \textsc{radmc-3d}. A recent detailed description of the radiative transfer method used here is presented in Appendix A in \cite{Franeck18} and further information can be found in the \textsc{radmc-3d} manual\footnote{\url{https://www.ita.uni-heidelberg.de/~dullemond/software/radmc-3d/manual_radmc3d/index.html}}. Furthermore, we include a thermal background in our post-processing, which is set to $T_{\rm bg}=2.73$ K \citep{Fixsen09}.
	
	The wavelengths at which the transitions occur are
	\mbox{$\lambda=2600.758\:\mathrm{\mu m}$} for the \mbox{CO(1-0)} transition and \mbox{$\lambda=2720.406\:\mathrm{\mu m}$} for the \mbox{{\COthird}(1-0)} transition. Unless stated otherwise, we always consider the J=$1\rightarrow0$ transition for {\CO} and {\COthird}, respectively. To take into account the turbulent motions within the gas leading to Doppler-shifted emission, we consider a velocity range of $\pm20$ km s$^{-1}$ over 201 channels, i.e. a channel width of $\Delta v = 0.2 \;{\rm km\; s}^{-1}$. We apply microturbulence which is equal to the sound speed in each cell. 
	
	\textsc{radmc-3d} also offers the option of so-called "tausurface" runs. These runs allow us to obtain the spatial location along the line of sight where the optical depth $ \tau \ge 1$. We use this option to identify the pixels were the line emission is optically thick (see Section \ref{sec:emission_analysis}).
	
	\subsection{Collisional partners}\label{sec:collpartner}
	\begin{figure*}
		\includegraphics[width=\textwidth]{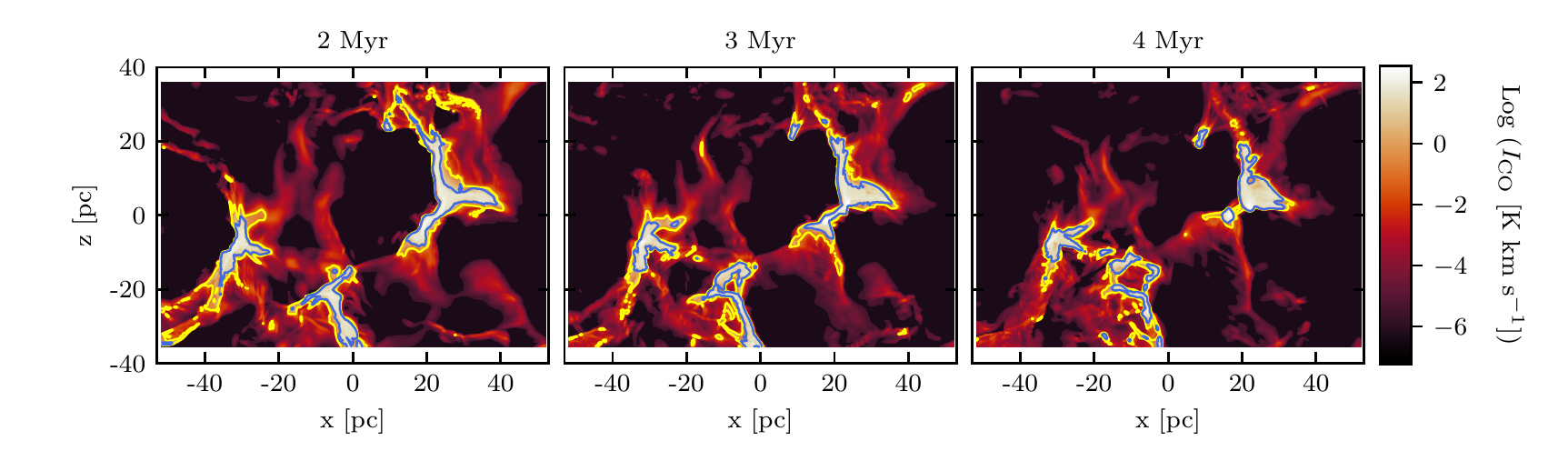}
		\caption{Synthetic emission maps showing the integrated intensity of the CO(1-0) transition 
		of \textit{MC2\_L10} in a projection along the $y-$direction. From left to right, the time evolution from 2 to 4 Myr can be followed. The yellow contour depicts the observable area of the cloud $A_{\mathrm{obs,CO}}$ and the blue contour outlines the region within which the CO(1-0) emission is optically thick. A significant part of the observable area is optically thick, and about 40 per cent of the cloud is CO-dark, with a density contrast of up to $\sim30$ between CO-dark and bright regions \citep[see Figures~2 and~8 of][]{Seifried20}.}
		\label{fig:mom0_times}
	\end{figure*}
	
	Collisional excitation determines the level populations in non-LTE conditions. Previous authors \citep[e.g.][and many more]{Seifried17b, Seifried20, Clarke18, Gong18, Szuecs14, Szuecs16} have only considered collisions between CO and para-/ortho-{\molh} \citep{Yang10}, as provided by the Leiden Atomic and Molecular Database (LAMDA) \citep{Schoier05}. Through the use of BASECOL \citep{basecol}, which is part of the Virtual Atomic and Molecular Data Centre (VAMDC)\footnote{\url{https://basecol.vamdc.eu/}}, we also obtained the collision rates for CO with Helium \citep{Cecchi02}. \cite{Walker15} furthermore provided the collision rates for CO with atomic hydrogen which we obtained from the BullDog Database\footnote{\url{https://www.physast.uga.edu/amdbs/excitation/CO/index.html}}. The collision rates for CO are presented in Fig.~\ref{fig:collision_rates} in Appendix \ref{app:collision_rates}. BASECOL and the BullDog Database do not provide the additional collision partners for {\COthird}. As the collision rates for {\COthird} with {\molh} obtained from the LAMDA database are identical to the ones provided for CO, we assume that the same holds true for the collision rates of {\COthird} with He and H.
	 
	Helium is not explicitly included in our chemical network. Instead, a fixed He abundance of \mbox{$\chi_{\mathrm{He}}=0.1$} \citep{Sembach00} with respect to the total number of hydrogen nuclei is assumed. As He is a noble gas it rarely forms molecules. On the other hand, we do not consider nearby O-stars such that Helium is rarely ionized. Hence, most of the Helium is neutral in the environmental conditions we consider. Hence, the He number density can easily be calculated for each grid cell in the SILCC-Zoom simulations and is provided to \textsc{radmc-3d} in the same format as the number densities of all other chemical species.

	\subsection{Emission maps}
	We compute the synthetic emission maps of the CO(1-0) and {\COthird}(1-0) transition using \textsc{radmc-3d} for all the simulations listed in Table \ref{tab:refinement} and for different evolutionary times, 1.5, 2, 2.5, 3 and 4 Myr after $t_0$ (see Section~\ref{sec:SILCC}). Further, for each simulation and each time, we vary the line-of-sight to point either along the $x-$, $y-$, or $z-$direction (i.e. we calculate 396 \textsc{radmc-3d} simulaitons in total). The size of the pixel of the resulting emission maps from \textsc{radmc-3d} is identical to the maximum refinement level from Table \ref{tab:refinement}. 
	
	For the analysis of the emission maps, we define an observational limit where the integrated intensity is \mbox{$ \ge 0.1\:\mathrm{K\: km/s} $}. On a given map, the area within which this criterion is fulfilled is called $A_{\mathrm{obs,CO}}$ for CO and $A_{\mathrm{obs,^{13}CO}}$ for {\COthird}.
	For example, Fig.~\ref{fig:mom0_times} shows the time evolution of CO emission maps obtained for the run \textit{MC2\_L10} projected along the y-direction. The emission maps depict the integrated intensity, $I_{\rm CO}$, in units of [${\rm K\; km\; s}^{-1}$]\footnote{Note that $I_{\rm CO}$ can be equally expressed in units of [erg (s sr cm$^2$)$^{-1}$], which is necessary when calculating the total luminosity.}, which is obtained by integrating the brightness temperature over all available velocity channels:
	\begin{equation}
	    I_{\rm CO} = \int T_{\rm B, CO}dv.
	\end{equation}
	Due to the inclusion of a thermal background in \textsc{radmc-3d} we afterwards deduct the background from our emission map before continuing with our analysis.
	
	A yellow contour indicates the observable area $A_{\mathrm{obs,CO}}$, while the blue contour indicates where the cloud is optically thick according to the results from the \textsc{radmc-3d} "tausurface" run (see above).
	
    \subsection{Total luminosity and impact of the collisional partners}
	\begin{figure}
		\includegraphics[width=\columnwidth]{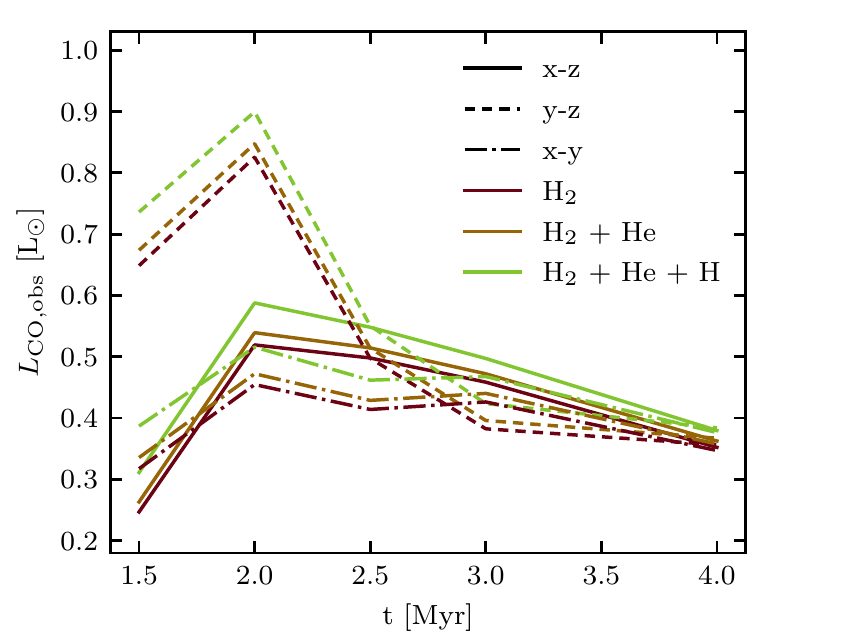}
		\caption{Total luminosity of CO(1-0) in the observable area $A_{\mathrm{obs,CO}}$ for the radiative transfer calculations with different collision partners. Shown are the luminosities of \textit{MC2\_L10} at different times. For all cases we use the same area in our calculations, the observable area $A_{\mathrm{obs,CO}}$ of the simulations using only {\molh} as collision partners. The dark red lines show the results using just {\molh} as collision partners, as provided in the molecular file from LAMDA. Furthermore, the figure shows the luminosity when additionally including He (orange lines) and He plus H (green lines) alongside {\molh} as collision partners. The results show that the addition of He as a collision partner increases the luminosity slightly by $\sim2-6 $ per cent, while the increase for CO is about $\sim7-26 $ per cent when additionally considering H.}
		\label{fig:lum_col_partners}
	\end{figure}

    It is useful to calculate the total luminosity of an emission map in order to compare between different simulations. The total luminosity, $L_{\rm CO, obs}$ is given as
    \begin{equation}
        L_{\rm CO, obs} = 4\pi d^2 F_{\rm CO, obs}\, ,
        \label{eq:lum}
    \end{equation}
    where we choose an arbitrary distance $d$ (as the distance will cancel out under the assumption that the side length of the pixel, $a$, fulfils $a<<d$). $F_{\rm CO, obs}$ is the total flux derived from the integrated intensity map, by adding up the contributions from the total number of pixels within the observable area, $n_{\rm obs}$:
    \begin{equation}
        F_{\rm CO, obs} = \sum_{i \,\in\, n_{\rm obs}} I_{\rm CO,i} A_{\rm pixel,i}.
    \end{equation}
    $A_{\rm pixel,i}$ is the size of square pixel $i$ in steradians given as
    \begin{equation}
        A_{\rm pixel,i}=\left(\tan^{-1}\left(\frac{a}{d}\right)\right)^2.
    \end{equation}
    In each of our synthetic maps all pixels have the same size $a$, given by the cell size on the current maximum refinement level of the 3D simulation (see above).

    We investigate the impact of using different collisional partners on the total luminosity. Therefore, we calculate different \textsc{radmc-3d} models. In the simplest case we only use {\molh} as the only collisional partner (following previous works).
	Secondly, we also include He and, thirdly, He plus H are considered in addition to {\molh} as collisional partners. Fig.~\ref{fig:lum_col_partners} shows $L_{\rm CO, obs}$ as a function of time resulting from the different sets of considered collisional partners for model \textit{MC2\_L10} projected along different lines of sight (LOS). For all three cases we use the same area when calculating the luminosities, the observable area $A_{\mathrm{obs,CO}}$ from the simulation using only {\molh} as a collision partner. When only considering collisions with {\molh} (dark red lines), we obtain the lowest $L_{\rm CO, obs}$. Including Helium (orange lines) increases the total luminosity by $\sim2-6 $ per cent, while including also atomic hydrogen (green lines) raises $L_{\rm CO, obs}$ by $\sim7-26 $ per cent for any time and LOS. Including H and He mostly enhances the integrated intensity near the boundary area of the observable region (see Fig.~\ref{fig:ratio_collision} in Appendix \ref{app:collision}, which shows the ratio of the integrated intensity including and excluding the additional collisional partners). Fig.~\ref{fig:ratio_hist} in Appendix~\ref{app:collision} shows a 2D PDF of the ratio of the integrated intensity maps including He and H and the maps excluding them against the integrated intensity including all collision partners. The PDF shows that the intensity changes in the most diffuse and the most dense regions of the cloud. Considering the significant impact which the additional collisional partners have on the synthetic emission maps, we include all of them in the following analysis.

	%%%%%%%%%%%%%%%%%%%% COLUMN DENSITY CO %%%%%%%%%%%%%%%%%%
	
	\section{CO column density}\label{sec:column_density}
	
	\begin{figure}
	    \centering
	    \includegraphics[width=\columnwidth]{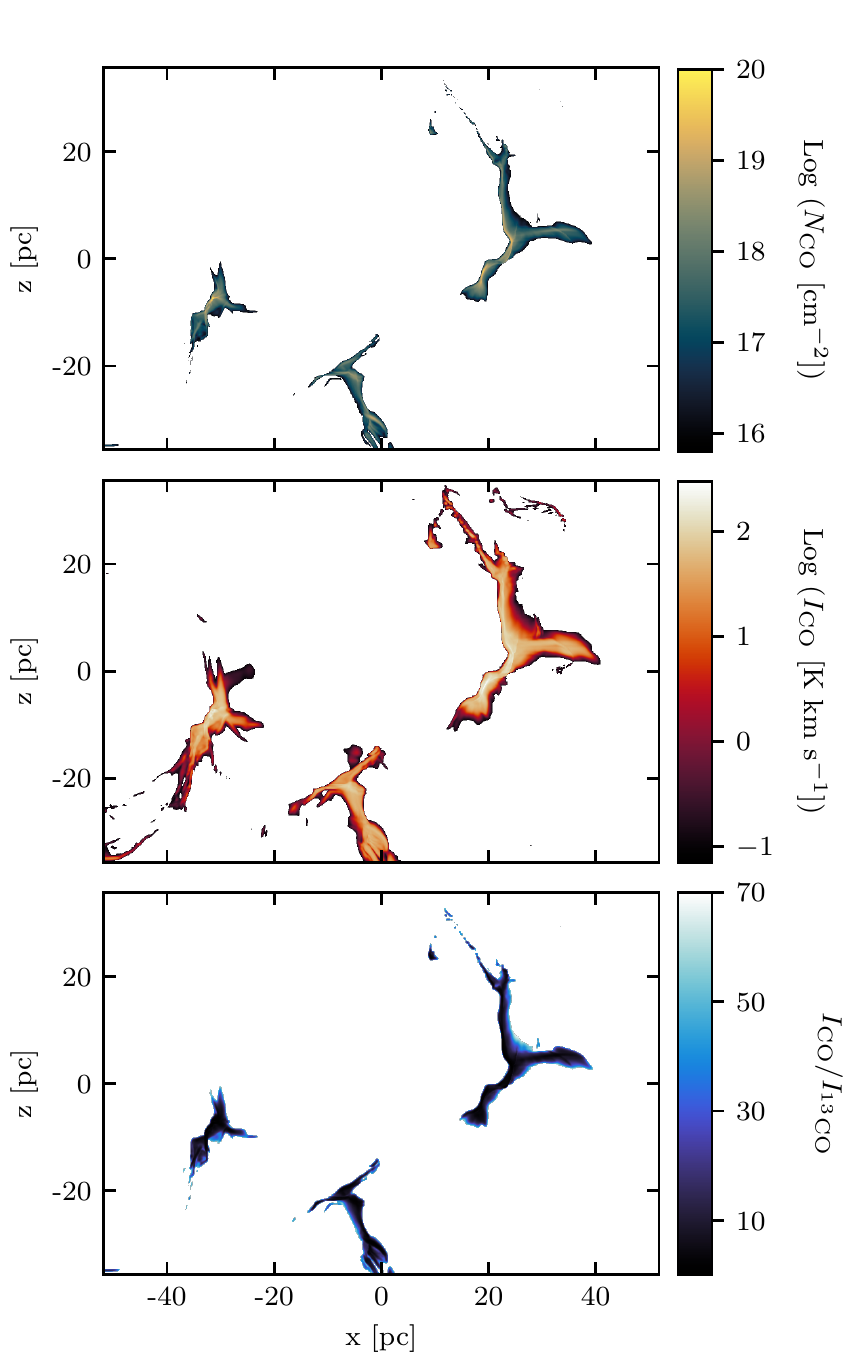}
	    \caption{{\it Top:} the column density $N_{\mathrm{CO}}$ for the run \textit{MC2\_L10} at 2 Myr as calculated directly from the simulation data within $A_{\mathrm{obs,^{13}CO}}$. {\it Middle:} the integrated intensity of the CO(1-0) transition within the observable area of this line, $A_{\mathrm{obs,CO}}$. {\it Bottom:} the ratio of the emission maps from CO(1-0) within the area where both tracers are observable, i.e. within $A_{\mathrm{obs,^{13}CO}}$. While a ratio of 69 is expected, the ratio becomes as low as 2 - 5 due to opacity effects, in agreement with observational results (see e.g. \citealt{Pineda08}; \citealt{Kong18} and others). All plots show the maps along the $y-$direction.}
	    \label{fig:ratio-MC2}
	\end{figure}
	
	\begin{figure}
		\includegraphics[width=1\columnwidth]{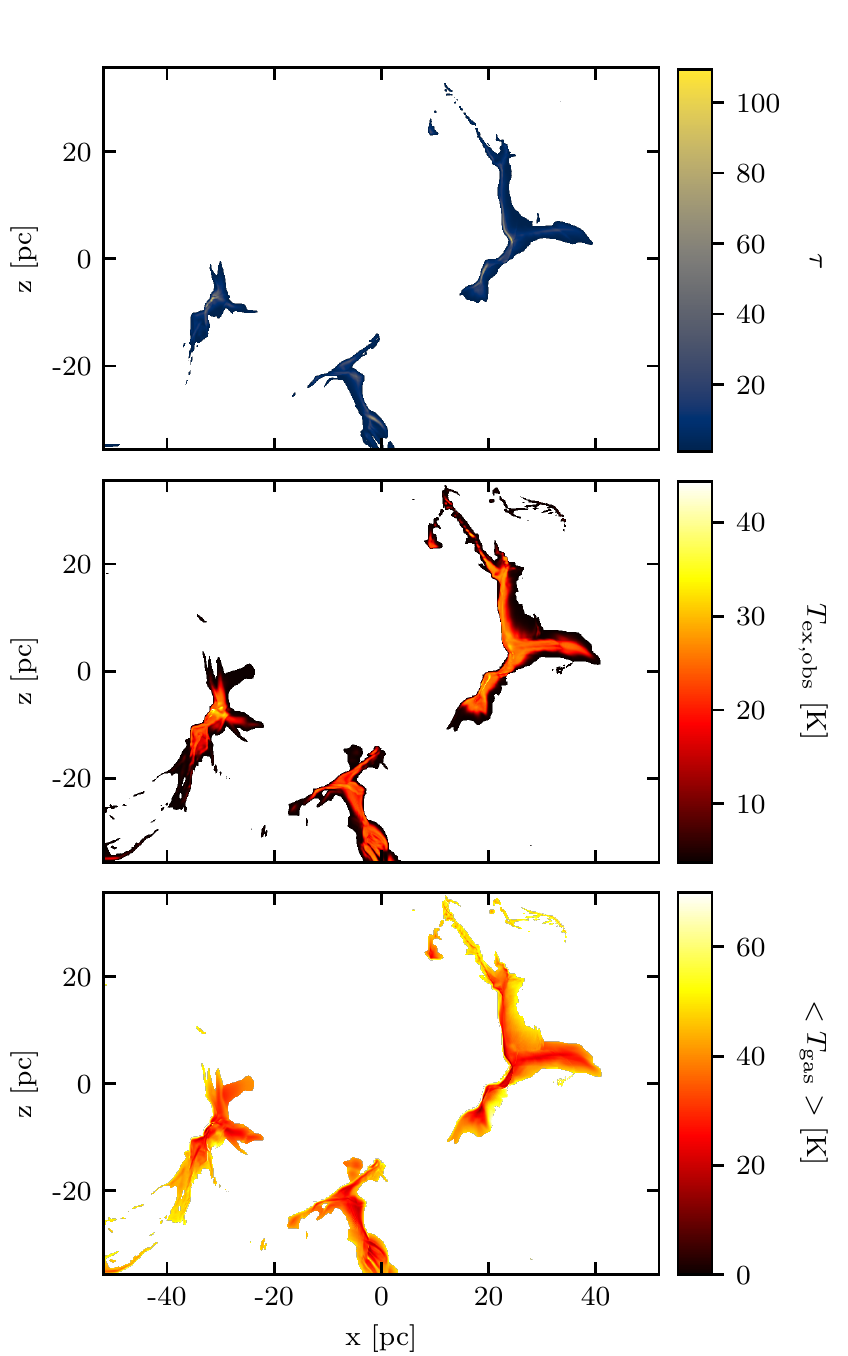}	
		\caption{Map of $\tau$ of {\CO}(1-0) (top) from solving Equation \eqref{eq:tau} and map of $T_{\mathrm{ex,obs}}$ (middle) from Equation \eqref{eq:Tex_obs} for {\it L10\_MC2} at 2 Myr for the $x-z$ projection (as in Fig.~\ref{fig:ratio-MC2}). The bottom panel shows the mass-weighted gas temperature. Only the observable region of the cloud is shown: the top panel shows the region for $A_{\mathrm{obs,^{13}CO}}$ as $\tau$ can only be estimated when both CO and {\COthird} are present; the middle and bottom panels show the regions for $A_{\mathrm{obs,^{12}CO}}$. The maximum excitation temperature is \mbox{44.4 K} while the intensity-weighted average of the excitation temperature map is \mbox{22.7 K}. The map of the gas temperature illustrates an inverse relation to the excitation temperature within the molecular cloud.}
		\label{fig:T_ex_obs}
	\end{figure}
	
	When observing regions of interest using line emission from different molecules, the column density (of the observed molecule) is usually unknown, unless otherwise determined (e.g. indirectly by using a different tracer, for example dust). In order to determine the column density $N$ for a certain molecule, the following equation is used \citep{Seifried17a}:
	\begin{equation}		
	N=\frac{8\pi \nu^3}{c^3}\frac{1}{A}f(T_{\mathrm{ex}})\frac{Q}{g_u}\frac{k_B/(h\nu)}{f(T_{\mathrm{ex}})-f(T_{\mathrm{bg}})}\frac{\tau}{1-e^{-\tau}}e^{\frac{E_u}{T_{\mathrm{ex}}}}\int T_Bd\nu.
	\label{eq:N_CO}
	\end{equation}
	Here, $\nu$ is the frequency and $A$ is the Einstein coefficient of the observed, spontaneous line transition of the molecule; $g_u$ is the degeneracy and $E_u$ the energy of the upper level in Kelvin; $h$ is the Planck constant, $k_B$ the Boltzmann constant and $c$ is the speed of light. $T_B$ is the brightness temperature obtained from the emission maps, $T_{\mathrm{bg}}$ is the background temperature and $\tau$ is the mean optical depth. The partition function $Q$ depends on the excitation temperature $T_{\mathrm{ex}}$. For a linear rotor, such as CO, $Q$ is approximated as
	\begin{equation}
	Q = \frac{k_B T_{\mathrm{ex}}}{hB} + \frac{1}{3} + \frac{1}{15}\frac{hB}{k_B T_{\mathrm{ex}}} + ... \approx \frac{k_B T_{\mathrm{ex}}}{hB},
	\end{equation}
	with the rotational molecular constant $B=57635.968\:{\rm MHz}$ (obtained form the Jet Propulsion Laboratory (JPL) Molecular Spectroscopy database and spectral line catalog; \citealt{Pickett98}).\footnote{Online at \url{ http://spec.jpl.nasa.gov} in the Catalogue directory.}
	
	$f(T_{\mathrm{ex}})$ is given by
	\begin{equation}
	f(T_{\mathrm{ex}}) = \frac{1}{\exp \left(\frac{h\nu}{k_B T_{\mathrm{ex}}}\right)-1}.
	\end{equation}
	
	If the observed molecule is optically thin, then Eq.~\eqref{eq:N_CO} can be used with the exclusion of the optical depth term $\frac{\tau}{1-e^{-\tau}}$, as it approaches unity. As CO is optically thick, an additional molecule that is optically thin needs to be observed, for which the isotopic ratio to the desired molecule is known and it is assumed that they trace the same gas. Observations of both these molecules are used to determine $\tau$. For CO this is assumed to be possible with the \mbox{CO(1-0)} and \mbox{{\COthird}(1-0)} transitions. Using the brightness temperatures $ T_{B, \mathrm{CO}} $ and $ T_{B, \mathrm{^{13}CO}} $, the mean optical depth $\tau$ can be obtained from solving a relation for the integrated brightness temperatures (e.g. \citealt{Arzoumanian13})
	\begin{equation}
	\frac{\int T_{B,\mathrm{CO}}d\nu}{\int T_{B, \mathrm{^{13}CO}}d\nu}=\frac{1-e^{-\tau_{\mathrm{CO}}}}{1-e^{-R\tau_{\mathrm{CO}}}},
	\label{eq:tau}
	\end{equation}
     using the Newton root finding algorithm. We assume the simple isotopic ratio $R=69$ \citep{Wilson99}.
     
	In the top panel of Fig.~\ref{fig:ratio-MC2} we plot the CO column density, $N_{\mathrm{CO}}$, which we directly obtain from the simulation data of run \textit{MC2\_L10} at 2 Myr, in the observable area $A_{\mathrm{obs,^{13}CO}}$, which is derived from the \textsc{radmc-3d} run. The CO column density is the one we wish to estimate using Eq.~\eqref{eq:N_CO}. The CO column density clearly increases towards the centres of the observable areas. For the observable area we use $A_{\mathrm{obs,^{13}CO}}$ in this case, because the \mbox{{\COthird}(1-0)} line is dimmer than \mbox{CO(1-0)} and therefore the observability of {\COthird} is the more restrictive condition. The observable region for the CO(1-0) transition, $A_{\mathrm{obs,CO}}$, is more extended. This can be seen in the middle panel of Fig.~\ref{fig:ratio-MC2}, which shows the synthetic emission map of the CO(1-0) transition within $A_{\mathrm{obs,CO}}$. The bottom panel depicts the ratio of the integrated intensities for \mbox{CO(1-0)} and \mbox{{\COthird}(1-0)} within the area where both lines are observable. In the areas with lower column density, the ratio $I_{\mathrm{CO}}/I_{\mathrm{^{13}CO}}$ is close to 69, which corresponds to the isotopic ratio used to estimate {\COthird}, while the ratio drops to values as low as 2-5 in the dense regions of the clouds due to opacity effects. This finding is in agreement with observational results (see e.g. \citealt{Pineda08}; \citealt{Kong18}). 
	
	The first step to determine $N_{\rm CO}$ from Eq. \eqref{eq:N_CO} is to estimate the opacity of the CO line, $\tau_{\rm}$ which is done using Eq. \eqref{eq:N_CO}. The top panel of Fig.~\ref{fig:T_ex_obs} depicts the computed $\tau_{\rm CO}$ for \textit{MC2\_L10} at 2 Myr projected along the $y-$direction. The \mbox{CO(1-0)} line is correctly determined to be optically thick in a large part of the observable area, as previously shown in Fig.~\ref{fig:mom0_times}.

	%%%%%%%%%%%%%%%%%%%
	\subsection{Different ways to determine the excitation temperature}
	
	    \begin{figure}
		\centering
		\includegraphics[width=\columnwidth]{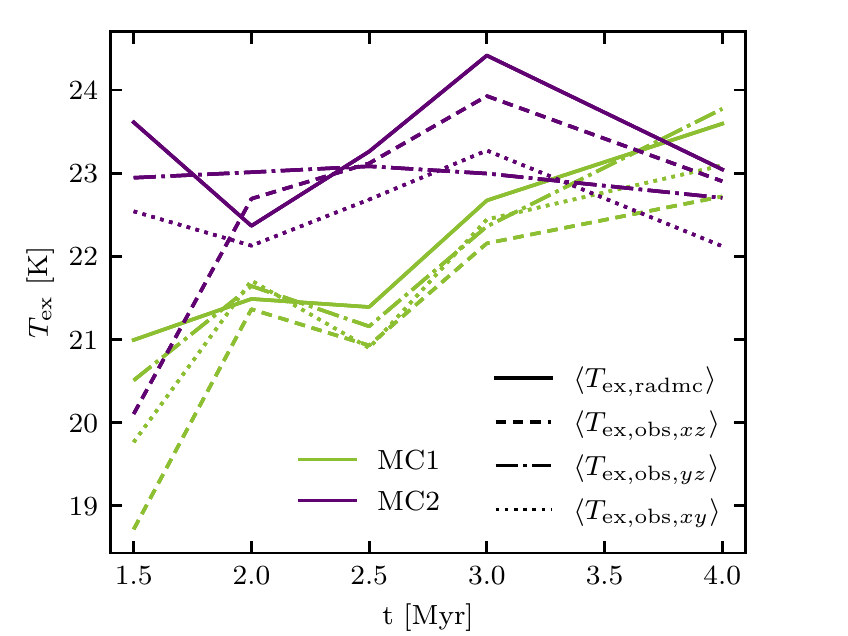}
		\caption{Time evolution of the mass-weighted excitation temperature of the map from \textsc{radmc-3d}, $\langle T_{\mathrm{ex,radmc}}\rangle$, following Eq. \eqref{eq:Tex_radmc_av} as well as the intensity-weighted observational results, $\langle T_{\mathrm{ex,obs}}\rangle$, following Eq. \eqref{eq:Tex_obs_av} for the three different projections along the principal coordinate axes (see test for further explanations). The excitation temperature varies by about 10 per cent ($\sim20$ per cent for \mbox{1.5 Myr}).}
		\label{fig:T_ex_t}
	    \end{figure}
	    
	Apart from $\tau_{\rm CO}$, the excitation temperature, $T_{\mathrm{ex}}$, is required in order to calculate $N_{\rm CO}$. There are several ways to determine $T_{\mathrm{ex}}$. The simplest approach is to assume a constant excitation temperature for the whole cloud. When using this approach we will set $T_{\mathrm{ex,fix}}=15$ K, which is a typical value for cold molecular clouds \citep{Stahler}, and which is comparable to the mass averaged gas temperature within our simulations (see bottom panel in Figure~\ref{fig:T_ex_obs}). We choose this as a fixed value as in thermal equilibrium the gas temperature should be equal to the excitation temperature.
	
	A more sophisticated approach is to determine the excitation temperature $T_{\mathrm{ex,obs}}$ from the emission of the observed molecule. There are two possibilities to do so: (i) if different transition lines from the same molecule would are available,  $T_{\mathrm{ex,obs}}$ could be obtained from population diagrams which are fitted with the Boltzmann distribution \citep{Goldsmith99}; (ii) otherwise the peak radiation temperature along the LOS can be used to estimate the excitation temperature in each pixel. Here, we adopt the second approach.
    Following \cite{Pineda08}, the radiation temperature is given as
	\begin{equation}
	T_R=T_0\left(\frac{1}{e^{T_0/T_{\mathrm{ex}}}-1}-\frac{1}{e^{T_0/T_{\mathrm{bg}}}-1}\right)\left(1-e^{-\tau}\right),
	\label{eq:rad_temp}
	\end{equation}
	with $T_0=h\nu(^{12}{\rm CO(1-0)})/k_{\rm B}=5.5$ K.
	The background temperature is set to \mbox{$T_{\mathrm{bg}}=2.73$ K}. The CO(1-0) line is optically thick, i.e. we assume that $\tau\rightarrow \infty$ and therefore \mbox{$(1-e^{-\tau})\rightarrow1$}. 
	For each pixel, we set the radiation temperature equal to the peak brightness temperature in velocity space $T_{\rm R} = T_{\mathrm{R,max}}(\mathrm{CO})$. We can now rearrange Eq.~\eqref{eq:rad_temp} to calculate the excitation temperature in each pixel as
	\begin{equation}
	T_{\mathrm{ex,obs}}=\frac{5.5~\mathrm{K}}{\ln\left\lbrace1+5.5~\mathrm{K}\times \left(T_{\mathrm{R,max}}(\mathrm{CO}) + 0.82~\mathrm{K}\right)^{-1}\right\rbrace }.
	\label{eq:Tex_obs}
	\end{equation}
	The intensity-weighted average of the whole map is calculated using the integrated intensity $I_{\rm CO}$ excitation temperature $T_{\rm ex,obs}$ and of each pixel as follows
	\begin{equation}
	    \langle T_{\rm ex,obs} \rangle= \frac{\sum\limits_{i \in A_{\rm obs,CO}} T_{\rm ex,obs} I_{\rm CO}}{\sum\limits_{i \in A_{\rm obs,CO}} I_{\rm CO,i}}.
	    \label{eq:Tex_obs_av}
	\end{equation}
	
	In Fig.~\ref{fig:T_ex_obs} (middle panel), we show a map of $T_{\mathrm{ex,obs}}$ within $A_{\mathrm{obs,CO}}$. The maximum excitation temperature of this map is at $T_{\mathrm{ex,obs,max}}=44.4$ K, while the intensity-weighted average excitation temperature is only $\langle T_{\mathrm{ex,obs}}\rangle=22.7$ K. The bottom panel shows the mass-weighted gas temperature in our simulation with a minimum gas temperature of \mbox{13 K} and a mass-averaged gas temperature of $\sim15$ K across our simulations. The gas temperature plot illustrates how the excitation temperature and the gas temperature show an inverse trend within the cloud. 

    Lastly, we may directly extract the excitation temperature from the \textsc{radmc-3d} level population calculation. For this we use the densities of each level as well as the Boltzmann distribution. The level population follows the relation
	\begin{equation}
	\frac{n_l}{n_u}=\frac{g_l}{g_u}e^{\frac{h\nu}{k_\mathrm{B} T_{\mathrm{ex}}}}, 
	\end{equation}
	with $n_u$ and $n_l$ the densities of the upper and lower level, respectively, which are obtained from the level population output file and $g_u=3$ and $g_l=1$ being the statistical weights of the upper and lower level. The line frequency for the \mbox{CO(1-0)} transition is $\nu=115.27\:{\rm GHz}$.
	Rearranging this equation gives the excitation temperature, $T_{\mathrm{ex,radmc,ijk}}$ within each grid cell (identified using cell indices $i,j,k$) as
    \begin{equation}
        T_{\mathrm{ex,radmc,ijk}}=-\frac{h\nu}{k_\mathrm{B}} \left[\ln\left(\frac{n_u}{n_l}\frac{g_l}{g_u}\right)\right]^{-1}.
        \label{eq:Tex}
    \end{equation}
    Since the output is in the 3D cube of the simulation, we calculate the mass-weighted average along the LOS for each pixel in 2D to create a map of the excitation temperatures $T_{\mathrm{ex,radmc,ij}}$ using
    
    \begin{equation}
        T_{\rm ex,radmc,ij} = \dfrac{\sum\limits_{k} m_{\rm CO,ijk} T_{\mathrm{ex,radmc,ijk}}}{\sum\limits_{k} m_{\rm CO,ijk}}
    \end{equation}
    Here $m_{\rm CO,ijk}$ is the CO mass in each cell as obtained from the 3D simulation data.
    Then, $T_{\mathrm{ex,radmc,ij}}$ is used to calculate the corresponding column density of pixel ($i,j$).
    
    For the mass-weighted excitation temperature averaged over the entire map, $\langle T_{\mathrm{ex,radmc}}\rangle$, we calculate
    
    \begin{equation}
        \langle T_{\mathrm{ex,radmc}}\rangle=\dfrac{\sum\limits_{pix \in A_{\rm obs,CO}} n_{\rm CO,pix}T_{\rm ex,radmc,pix}}{\sum\limits_{pix \in A_{\rm obs,CO}} n_{\rm CO,pix}} \, ,
        \label{eq:Tex_radmc_av}
    \end{equation}
    with the CO number density $n_{\rm CO}$ from the simulation, restricting ourselves to the observable area. We show the resulting time evolution $\langle T_{\mathrm{ex,radmc}}\rangle$ and the intensity-weighted excitation temperatures $\langle T_{\mathrm{ex,obs}}\rangle$ as obtained for the three different projections along the principal coordinate axes in Fig.~\ref{fig:T_ex_t}. Overall the observationally obtained average seems to be in good agreement with $\langle T_{\mathrm{ex,radmc}}\rangle$, deviating mostly with less than 10 per cent, only at \mbox{1.5 Myr} the deviation is at $\sim20$ per cent.
	
	\subsection{Deriving the CO column density based on different excitation temperature estimates}
	\begin{figure*}
		\includegraphics[width=\textwidth]{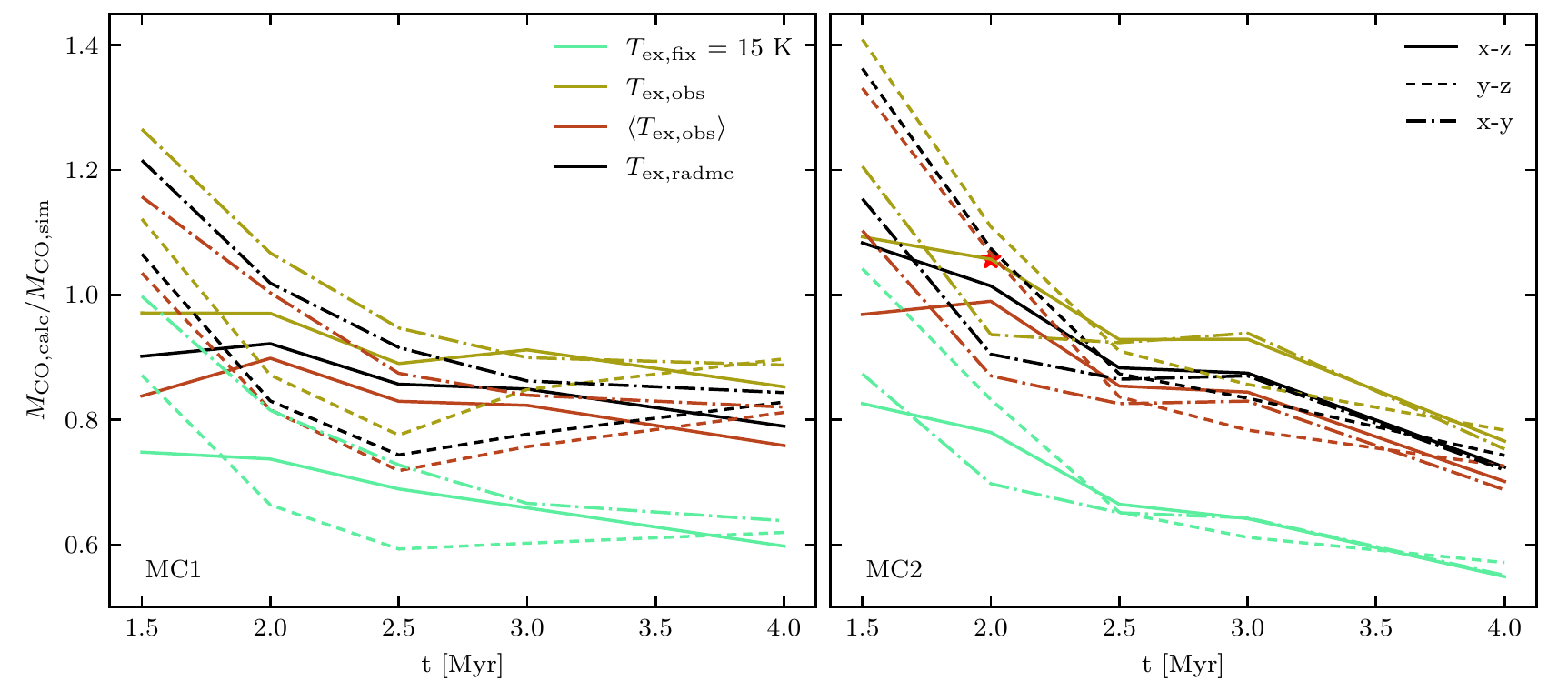}
		\caption{Ratio of the calculated (using Eq.~\eqref{eq:N_CO}) and the simulated CO masses (see Fig.~\ref{fig:M_co_h2}) for MC1 (left) and MC2 (right) as a function of time and for three different LOS. We show the results obtained using a fixed $T_{\rm ex,fix}=15$ K (similar to the average gas temperature; light green lines); the "observed" excitation temperature derived for each individual pixel, $T_{\rm ex, obs}$ (dark yellow lines); the intensity-weighted average of $T_{\rm ex, obs}$, i.e. $\langle T_{\rm ex, obs} \rangle$ (reddish lines); and the mass-weighted excitation temperature for each pixel obtained from the \textsc{radmc-3d} level population, $T_{\rm ex, radmc}$ (black lines). Overall the results show a downward trend as a function of time. The star marks the location of the results for $T_{\mathrm{ex,obs}}$ along the $y$-direction at 2 Myr, used for the 2D-PDF shown in Fig. \ref{fig:Nratio-N-MC2}.} 
		\label{fig:mass_obs}
	\end{figure*}
	\begin{figure}
		\includegraphics[width=\columnwidth]{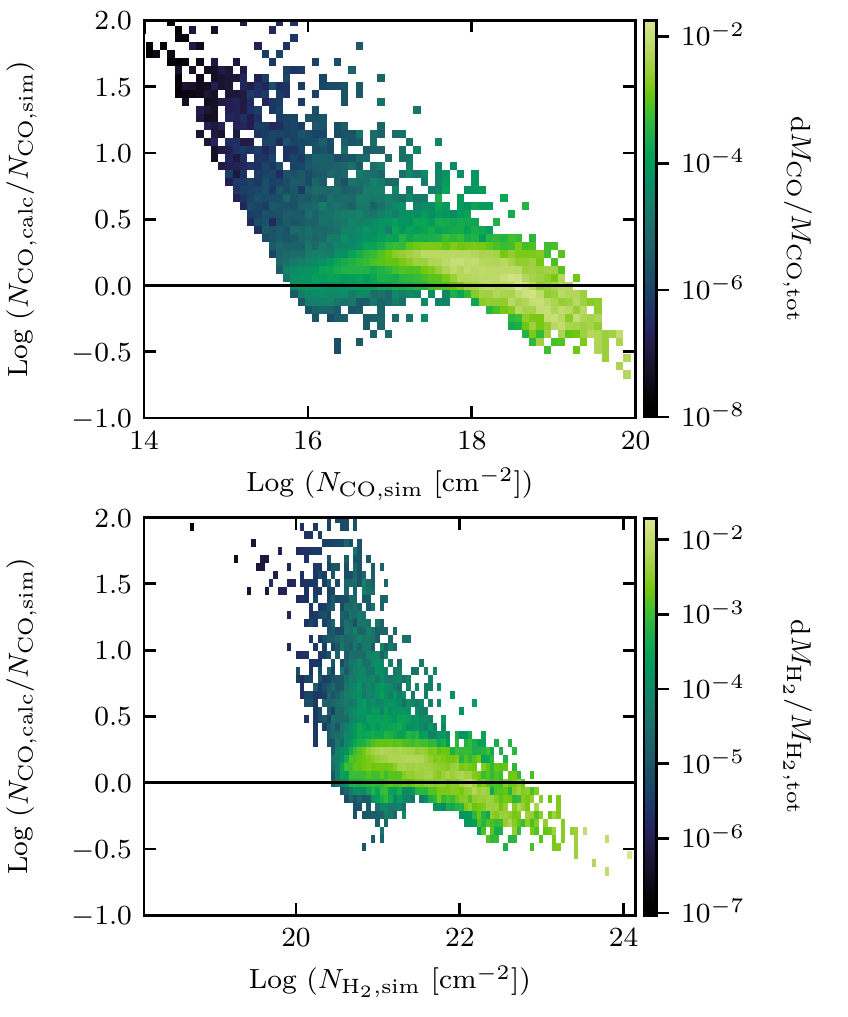}
		\caption{Two-dimensional PDF showing the ratio of $N_{\mathrm{CO,calc}}$ calculated from the synthetic emission map and $N_{\mathrm{CO,sim}}$ derived from the simulation data as a function of the simulation column densities of (i) CO (top), and (ii) {\molh} (bottom). The PDF is also mass-weighted in CO (top) and {\molh} (bottom), respectively. Shown are only pixels from the observable region for the $x-z$ projection of {\it MC2\_L10} at 2 Myr. Here, the case of $T_{\mathrm{ex,obs}}$ has been used to calculate $N_{\mathrm{CO,calc}}$. Especially in the high column density region, where most of the CO mass is located, $N_{\mathrm{CO,calc}}$ is much smaller than the actual CO column density because the intrinsic assumption that {\COthird} is optically thin breaks down. On the other hand, at low columns $N_{\mathrm{CO,calc}}$ is much larger than the true CO column density because both {\CO} and {\COthird} are optically thin. This suggests that different excitation temperature estimates are needed at low and high columns. Another thing to note is that a great portion of the {\molh} mass is not properly traced by CO, which can be explained by the presence of CO dark gas \citep[see][]{Seifried20}.}
		\label{fig:Nratio-N-MC2}
	\end{figure}

	After selecting an excitation temperature, we have everything we need in order to calculate the CO column density $N_{\mathrm{CO,calc}}$ following Eq.~\eqref{eq:N_CO}. From $N_{\mathrm{CO,calc}}$ we can easily get the CO mass, $M_{\mathrm{CO,calc}}$, by multiplying each $N_{\mathrm{CO,calc}}$ with the pixel area and summing over all pixels.
	We only consider pixels within the observable area of {\COthird}, i.e. all pixels within $A_{\mathrm{obs,^{13}CO}}$. 
	The calculated CO mass $M_{\mathrm{CO,calc}}$ can be directly compared with the total CO mass within the same area obtained from the CO density file of the simulation, $M_{\mathrm{CO,sim}}$.
	In Fig.~\ref{fig:mass_obs}, we show the ratio of $M_{\mathrm{CO,calc}}$ and $M_{\mathrm{CO,sim}}$ for four different excitation temperature estimates. These are (i) the fixed excitation temperature, $T_{\rm ex,fix}=15$ K (light green lines);(ii) the "observed" excitation temperature derived for each individual pixel, $T_{\rm ex,obs}$ from Eq.~\eqref{eq:Tex_obs} (dark yellow lines); the intensity-weighted average of $T_{\rm ex,obs}$, $\langle T_{\rm ex,obs}\rangle$ from Eq;~\eqref{eq:Tex_obs_av} (reddish lines); and the mass-weighted excitation temperature for each pixel from the \textsc{radmc-3d} level population, $T_{\rm ex,radmc}$ (black). 
	We show the results for {\it MC1\_L10} (left panel) and {\it MC2\_L10} (right panel) as a function of time for the three different LOS. The red star marks the calculation which is further discussed in Fig.~\ref{fig:Nratio-N-MC2} (see below).
    
    For both clouds and all excitation temperature estimates the CO mass ratio decreases as a function of time. At early times ($\sim$1.5 Myr), CO is still forming (see Fig.~\ref{fig:M_co_h2}) and is not yet optically thick as it is the case for later times, when a significant fraction of the CO-bright area is affected by optical depth effects (see Fig.~\ref{fig:mom0_times}). 
    For $T_{\mathrm{ex,fix}}=15~\mathrm{K}$, the CO mass is underestimated by $\sim20-45$ per cent after 2 Myr. We choose 15~K because this corresponds to the mass-averaged gas temperature of the CO-forming gas and would represent the excitation temperature in local thermal equilibrium (see bottom panel in Figure~\ref{fig:T_ex_obs}). However, Fig.~\ref{fig:mass_obs} suggests that $T_{\mathrm{ex,fix}}=15~\mathrm{K}$ might be too low and thus, we also calculate the CO mass ratio for different values of $T_{\mathrm{ex,fix}}=10,\,15,\,20,\,25,\,30$ K as shown in Fig.~\ref{fig:Nratio-t-MC2} in Appendix \ref{app:Tex}.
    We find that the CO mass ratio depends linearly on the assumed excitation temperature in this case. Despite the fact that the cloud is not yet forming stars, i.e. there is no feedback from within the cloud, a higher excitation temperature of $20-30$~K returns a more accurate estimate of the total CO mass than the average gas temperature of $15$~K. This finding is consistent with the intensity-weighted average excitation temperature of $\sim23$ K derived from Figs.~\ref{fig:T_ex_obs} and \ref{fig:T_ex_t}.
    
    On the other hand, using the pixel-wise derived excitation temperature, $T_{\rm ex,obs}$, leads to an overestimation of the CO mass of $\sim10-40$  per cent initially and an underestimation of $\sim10-25$ per cent later on, while $\langle T_{\rm ex,obs}\rangle$ (from Fig.~\ref{fig:T_ex_t}) and $T_{\rm ex,radmc}$ give similar results. 
    From the "tausurface" run from \textsc{radmc-3d} we derive that 20-55 per cent of the area used in the calculations is optically thick, resulting in up to 80 per cent being optically thin (shown in the top row in Fig~\ref{fig:deviations} in the next section). 
    In the optically thin regions the assumption made for the excitation temperature estimate does not hold true anymore. This leads to the overestimation of the CO mass. Simultaneously, we assume that the {\COthird}(1-0) line is optically thin when estimating $\tau_{\rm CO}$. However, our calculations show that this is not the case in the regions of highest {\COthird} emission. Hence, $\tau_{\rm CO}$, which is derived from the integrated brightness temperature ratio of CO/{\COthird} is underestimated, and in turn  $N_{\mathrm{CO,calc}}$ is  underestimated. As the latter occurs in the regions of the highest density where most of the mass is located, this deviation in the column density translates into a noticeable deviation in the total CO mass. 
	
	To understand in which regions of the cloud the CO column density is over or underestimated, we look at the mass-weighted, two-dimensional probability density function (PDF) of the $N_{\mathrm{CO,calc}}$/$N_{\mathrm{CO,sim}}$ ratio in Fig.~\ref{fig:Nratio-N-MC2} for the case of $T_{\rm ex,obs}$. The top panel shows the PDF over the present $N_{\rm CO,sim}$ in the simulation while the bottom panel shows the PDF over the corresponding $N_{\rm H_2,sim}$ in the simulation. We can clearly see that the column density is underestimated in regions of high CO column density, where most of the CO mass is located. At high CO columns {\COthird} is also optically thick and hence the $\tau_{\rm CO}$ we calculate is flawed. Simultaneously, in the regions where the CO column density is below \mbox{$\sim 10^{18}$ cm$^{-2}$}, $N_{\mathrm{CO,calc}}$ overestimates the true CO column density. For these columns {\CO} might be optically thin and hence, again, the assumption that one tracer is optically thin and the other is optically thick does no longer apply. Overall these results seem to be on par with observational results and variations (e.g. \citealt{Goldsmith08}). In order to further improve these calculations, additional CO lines could be used to calculate the excitation temperatures from multiple line ratios using the Boltzmann distribution. 
	
	Additionally to the over-/underestimation of $N_{\rm CO,calc}$, we can see in Fig.~\ref{fig:Nratio-N-MC2} that a great amount of the {\molh} mass is not properly traced by CO. This can be explained due to the presence of CO dark gas \citep[see][for a more detailed explanation]{Seifried20}. CO not properly tracing {\molh} as well as the over-/underestimation of $N_{\rm CO}$ at high/low columns make it difficult to directly use $N_{\rm CO}$ to find the correct $N_{\rm H_2}$. Possibly, new techniques involving machine-learning could improve the calculation of $N_{\rm CO}$.
    
    Overall, the intensity-weighted average excitation temperature $\langle T_{\rm ex,obs}\rangle$ from Eq.~\eqref{eq:Tex_obs_av} gives the best results for $N_{\rm CO}$. Typically, the errors in the total CO mass are of the order of $\sim$20 per cent in this case.
	
	%%%%%%%%%%%%%%%%%%%% SYNTHETIC EMISSION ANALYSIS %%%%%%%%%%%%%%%%%%
	\section{Impact of the numerical resolution on the synthetic CO emission and the {\xco} factor} \label{sec:emission_analysis}
	In this section we discuss how the properties of the synthetic CO emission maps depend on the resolution of the underlying 3D SILCC-Zoom simulation data.
	In particular, we will explore how the {\xco} factor (see Eq.\ref{eq:Xco}) changes with increasing effective resolution and whether the {\xco} factor is converged. \citet{Gong18} suggest that {\xco} is numerically converged at a resolution of $\sim$1~pc. This is in contrast to the findings of \citet{Seifried17b} and \citet{Joshi19}, who show that the CO abundance itself, and subsequently the CO mass and emission, is not converged at such coarse resolution. From these studies we suspect that also {\xco} will not converge at 1~pc, but that higher resolutions of the underlying 3D simulation data are required to derive the {\xco} factor from numerical simulations. 
	
	The numerical resolutions in this and the next section depict the resolution of the SILCC simulations. For post-processing with \textsc{radmc-3d} we set the resolution of the synthetic observations to the same value as the numerical resolutions. For our highest resolution runs this is much higher than actual observations are able to resolve. For comparison we have run synthetic observations on the same high numerical resolution simulations, setting the resolution in \textsc{radmc-3d} to lower levels. We found that this does not change the results in a significant way. Any changes are observed at the fifth or sixth significant digit. Any of the following results are therefore due to the numerical resolution used in the underlying 3D simulations.
	
	\subsection{Resolution study}
	\begin{figure}
		\centering
			\includegraphics[width=\columnwidth]{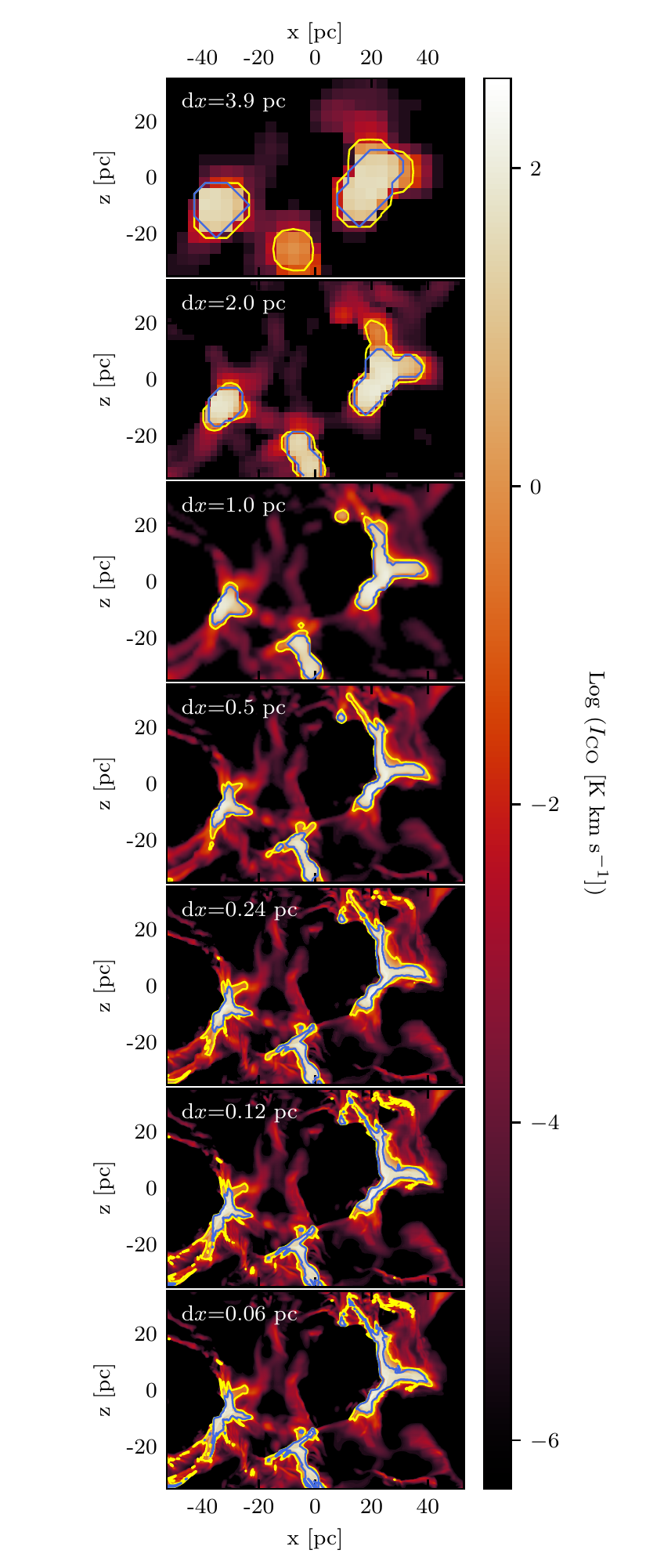}
			\caption{Synthetic emission maps showing the integrated intensity of the CO(1-0) transition for different spatial resolutions of the 3D SILCC-Zoom simulation data as indicated in the top left corner of each sub-map. Each map shows the front side of MC2 at 2 Myr. The yellow contour depicts the observable area of the cloud, $A_{\mathrm{obs,CO}}$, and the blue contour depicts the region where the cloud becomes optically thick according to the results of the "tausurface" run in \textsc{radmc-3d}. With increasing spatial resolution, the structure of the cloud changes from a blobby to a more defined, filamentary structure.}
			\label{fig:mom0_res}
		\end{figure}

	\begin{figure*}
		\includegraphics[width=\textwidth]{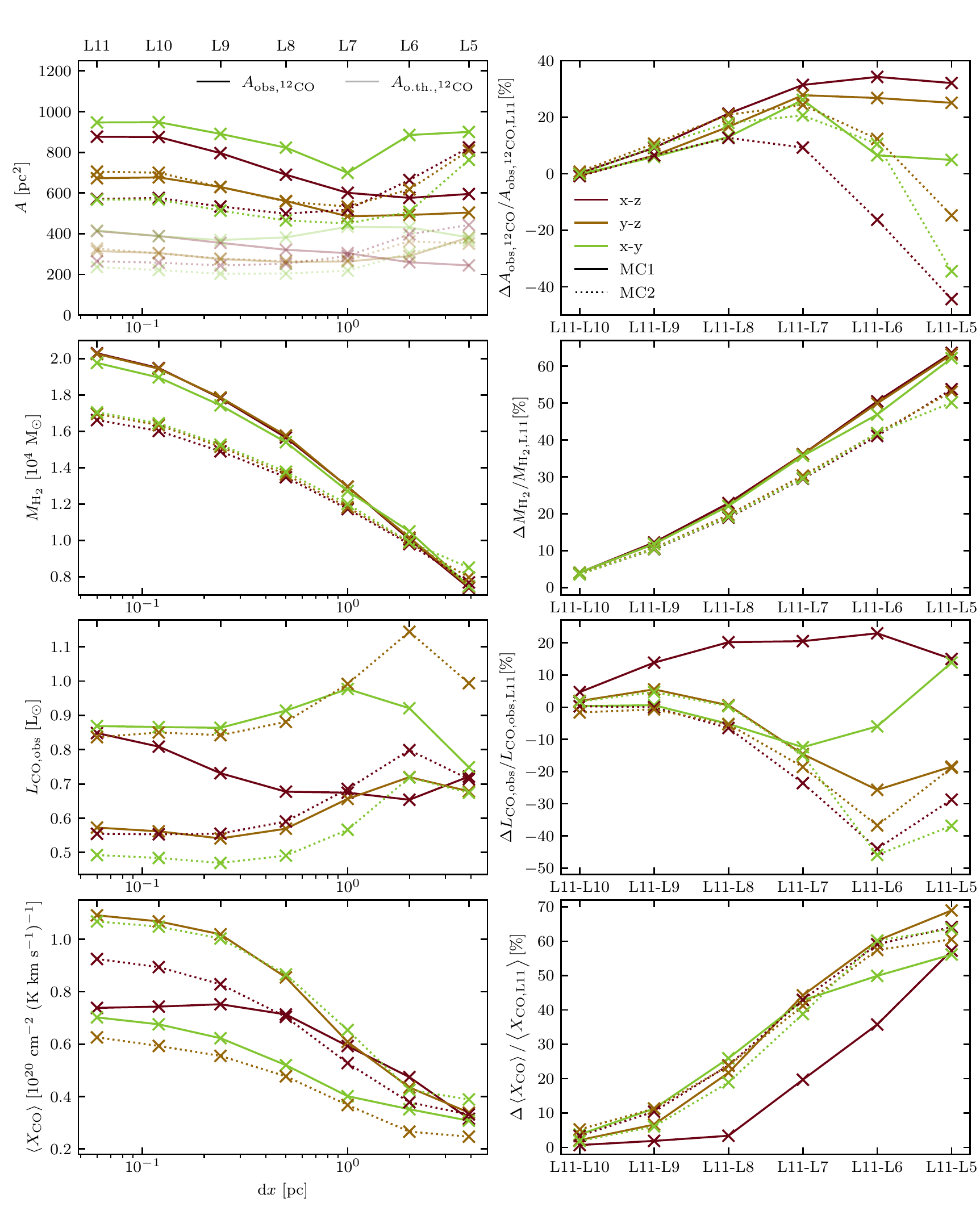}
		\caption{{\it Left column:} from top to bottom we show the observable and optically thick area, {\molh} mass in the observable area, the CO luminosity in the observable area as well as the resulting {\xco} factor for different spatial resolutions of the 3D SILCC-Zoom simulation data for both clouds at 2 Myr and for three LOS each as a function of the effective spatial resolution. {\it Right column:} the deviations between each resolution level and the highest level {\it L11} (corresponding to a maximum spatial resolution of d$x=0.06$ pc), given in per cent with respect to {\it L11}.}
		\label{fig:deviations}
	\end{figure*}

	As a first step, we present the $y-$projection of the \mbox{\CO(1-0)} integrated intensity maps of MC2 at $t=2$ Myr in Fig.~\ref{fig:mom0_res}. From top to bottom we increase the effective spatial resolution of the underlying SILCC-Zoom simulation data (as indicated in the top left corner of each sub-map). For a description of the different simulations see Table~\ref{tab:refinement} and Section~\ref{sec:SILCC}. The yellow contour outlines the observable region $A_{\mathrm{obs,CO}}$, while the blue contour indicates the area where CO is optically thick.
	
	The figure shows that with increasing spatial resolution, the structure of the cloud changes from a blobby to a more defined, filamentary structure. Furthermore, the maximum integrated intensity increases with increasing spatial resolution. The two contours approach each other with increasing resolution, showing that an increasingly large area of the brightest emission is optically thick. 
	The relative difference of the synthetic CO emission maps when increasing the resolution by a factor of 2 each is plotted in Figure~\ref{fig:lvl-diff-MC2} in Appendix \ref{app:differences}. The relative changes are decreasing for higher and higher resolutions, in particular in the parts where the cloud is optically thick.  
	
	To get a more quantitative understanding of the relative changes between the different spatial resolution levels, we present the relevant integrated quantities of the two clouds as a function of their spatial resolution in Fig.~\ref{fig:deviations} (MC1 with solid lines and MC2 with dotted lines). We again show the situation at $t=2$ Myr. The different colours show the different LOS. The left column depicts the derived quantities as a function of the spatial resolution, while the right column shows the deviation of the result obtained for a specific resolution level from the highest available spatial resolution ({\it L11}). From top to bottom we plot the observable area of the cloud $A_{\mathrm{obs,CO}}$ and the optically thick area $A_{\mathrm{o.th.,CO}}$ (pale lines), the {\molh} mass within the observable area (second row), the CO luminosity in the observable area (third row) and the resulting {\xco} factor (bottom row).
	
	With increasing spatial resolution, both clouds first show a decrease of the observable area, followed by an increase once the effective spatial resolution exceeds 1~pc and a stabilization of $A_{\mathrm{obs,CO}}$ below $\sim 0.125$~pc (note that the resolution increases from right to left). For instance, a pixel which is observable at {\it L5} is split into four smaller pixels at {\it L6}, from which only one might still be observable, leading to a decrease in the observable area.	Yet, the increase in the observable area is caused by the increase in the CO mass (see Fig.~\ref{fig:M_co_h2}) near the border of $A_{\mathrm{obs,CO}}$ (see Fig.~\ref{fig:lvl-diff-MC2} in Appendix \ref{app:differences}) for higher resolution. At 2 Myr little of the CO is frozen out and hence the increase is probably more pronounced at $t=$2~Myr than at later evolutionary times. The flattening of the curve shows that the observable area converges, and only changes about is $\sim \pm1 $ per cent when increasing the resolution from {\it L10} to {\it L11}, as seen in the right side of Fig.~\ref{fig:deviations}.
	The optically thick $A_{\mathrm{o.th.,CO}}$ seems to be some non-constant fraction of $A_{\mathrm{obs,CO}}$ and there is an increase in $A_{\mathrm{o.th.,CO}}$ from L10 to L11.
	
	The {\molh} mass within $A_{\mathrm{obs,CO}}$ shows the expected increase with increasing spatial resolution (compare with Fig.~\ref{fig:M_co_h2}). The small differences between the LOS stem from the slightly different $A_{\mathrm{obs,CO}}$ for each LOS, leading to small variations as a function of the chosen LOS. The {\molh} mass slowly converges with increasing resolution, with a difference between {\it L10} and {\it L11} of less than $\sim$5 per cent. Hence, even at an effective resolution of $<0.1$~pc the {\molh} mass is not fully converged at $t=2$~Myr.
	
	Surprisingly, there is no unique trend of decreasing or increasing CO luminosity with resolution level (third row). The different LOS show different trends for MC1, especially for an effective resolution coarser than d$x \gtrsim 0.25$ pc (corresponding to {\it L9}) but the relative changes drop below 10 per cent for all but one LOS. In case of MC2, all LOS show the same trend despite the different absolute values of the total luminosity. However, independent of how $L_{\rm CO,obs}$ evolves with decreasing d$x$, we see a clear trend of convergence with increasing d$x$. The remaining differences between {\it L10} and {\it L11} drop below $\sim \pm$5 per cent for both clouds and all LOS.
	
	%%%%%%%%%%%%%%%%%%%%%
	\subsection{{\avxco} factor}
	\begin{figure*}
		\includegraphics[width=\textwidth]{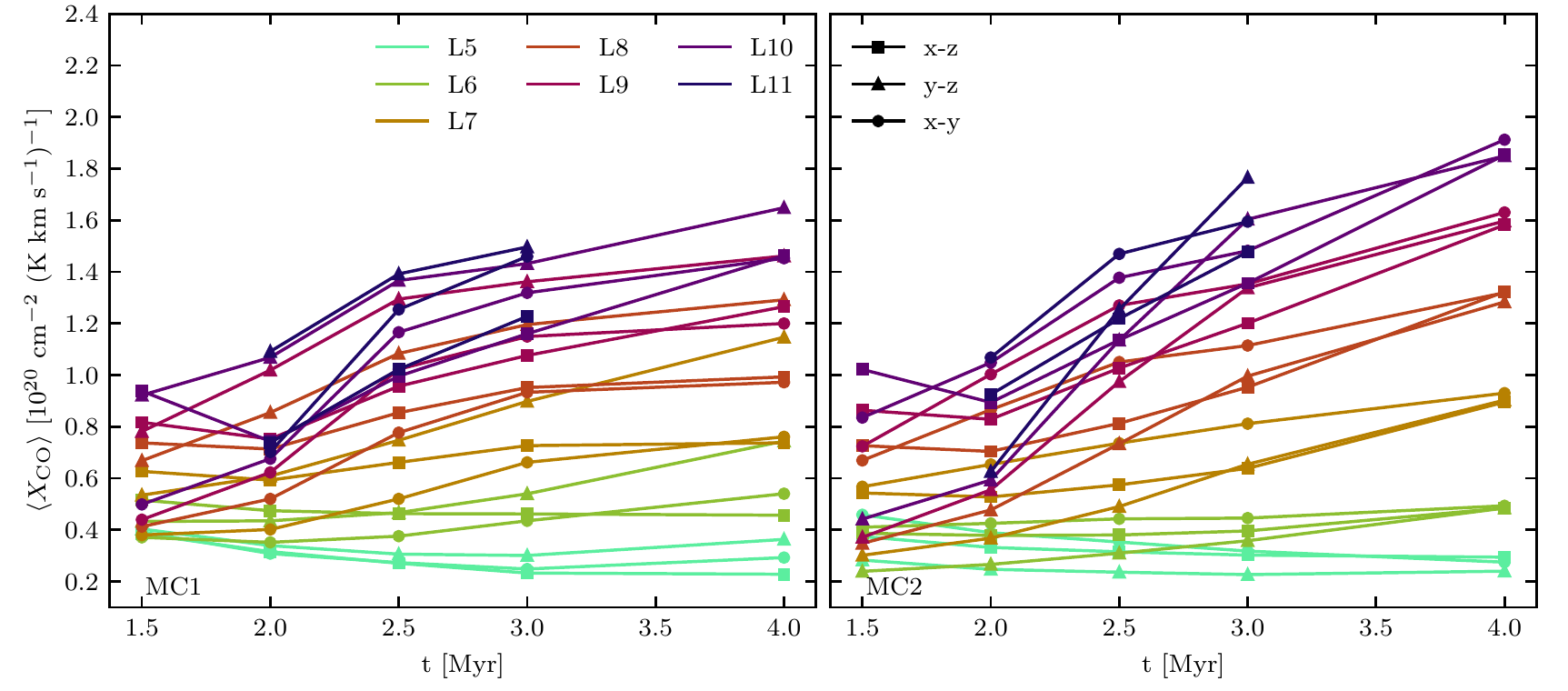}
		\caption{{\avxco} factor of the observable regions $A_{\mathrm{obs,CO}}$ for MC1 (left) and MC2 (right) over time. The different colours represent the different spatial resolutions in pc, while the symbols depict the different projections. Overall, the {\avxco} factor increases with increasing spatial resolution. Furthermore, for the results above d$x=2.0$ pc ({\it L6}), the factor also increases with time.}
		\label{fig:X_co}
	\end{figure*}
	
	We calculate the {\avxco} factor for each cloud using Eq.~\eqref{eq:Xco}, where we sum up the integrated quantities within all pixels in the observable region $A_{\mathrm{obs,CO}}$.
	In the bottom panels of Fig.~\ref{fig:deviations}, we show that the resulting {\avxco} increases with increasing spatial resolution from \mbox{$\sim 0.25-0.39\times 10^{20}\:{\rm cm}^{-2}\:({\rm K\: km\: s}^{-1})^{-1}$} at {\it L5} to \mbox{$\sim 0.63 - 1.09\times 10^{20}\:{\rm cm}^{-2}\:({\rm K\: km\: s}^{-1})^{-1}$} for {\it L11}. The increase in {\avxco} results from the increasing {\molh} mass, while $L_{\rm CO,obs}$ stays relatively flat. Similar to the underlying {\molh} mass and total CO luminosity, the relative difference between {\it L10} and {\it L11} decreases to below $\sim7$ per cent. A similar effect is seen for the other time steps in Figures \ref{fig:X_CO_analysis_MC1} and \ref{fig:X_CO_analysis_MC2} in Appendix \ref{app:xco}, where we plot the time evolution of {\avxco} as well as the underlying {\molh} mass and total CO luminosity. Since $L_{\rm CO,obs}$ is generally decreasing with increasing resolution, while $M_{\rm H_{2}}$ is increasing with resolution and with time, {\avxco} also increases with time and resolution. The deviations between {\it L10} and {\it L11} usually remain below $\sim 10$ per cent for all time steps. 
	
	This is also shown in Fig.~\ref{fig:X_co}, where we plot {\avxco} as a function of time for all simulated resolution levels, time steps, and LOS. In general we see that {\avxco} increases with increasing resolution. Even between {\it L10} and {\it L11} there are small remaining differences in {\avxco}. This indicates that a convergence in {\avxco} might be achieved only for simulations with an effective spatial resolution of d$x \lesssim 0.1$ pc. This is in disagreement with \cite{Gong18}, who suggest that {\avxco} is converged at $\sim 1-2$ pc, corresponding to our {\it L6} run (see Section~\ref{sec:Xco} for further discussions). The bottom right panel in Fig.~\ref{fig:deviations} shows that there is a $35-60$ per cent deviation between {\it L6} and {\it L11} for MC1 and a 60 per cent deviation for MC2. We further discuss the convergence with spatial resolution in the next section.
	
	Further, we note that for both clouds and any resolution analysed here, we do not reach the average Milky Way value of \mbox{$ X_{\mathrm{CO,MW}}=2\times 10^{20}\:{\rm cm}^{-2}\:({\rm K\: km\: s}^{-1})^{-1}$} \citep{Bolatto13}. For the analysed times and the highest resolution simulation data available for all times ({\it L10}), we obtain \mbox{$\left\langle X_{\mathrm{CO}}\right\rangle\approx 0.50-1.65~\times 10^{20}\:{\rm cm}^{-2}\:({\rm K\: km\: s}^{-1})^{-1}$} for MC1 and \mbox{$\left\langle X_{\mathrm{CO}}\right\rangle\approx 0.44-1.91~\times 10^{20}\:{\rm cm}^{-2}\:({\rm K\: km\: s}^{-1})^{-1}$} for MC2. In particular our young molecular clouds have \mbox{$ X_{\mathrm{CO}}\lesssim 10^{20}\:{\rm cm}^{-2}\:({\rm K\: km\: s}^{-1})^{-1}$}, so at least a factor of 2 smaller than $ X_{\mathrm{CO,MW}}$. 
    However, a smaller {\avxco} is not surprising because we only consider the CO luminosity and corresponding {\molh} within the observable area, while the Milky Way average naturally accounts for CO-dark molecular gas.

    Indeed, about 40 per cent of our clouds are CO-dark \citep[see Figures~2 and~8 in][]{Seifried20}. There is a density contrast of up to $\sim30$ between the CO-dark and CO-bright regions, with the number density for CO-dark gas being $\sim10-10^3\:{\rm cm^{-3}}$ and that for CO-bright gas being above $300\:{\rm cm^{-3}}$.
    Consequently, when the whole cloud (i.e. also the CO-dark {\molh}) is considered, {\avxco} can even be larger than the Milky Way average.% as discussed in \citet{Seifried20}. 
    We show this in Fig.~\ref{fig:X_co_of_map} in Appendix \ref{app:x_co_of_map}, where we recalculate {\avxco} for the whole map rather than the observable area only. Including the CO-dark gas, our {\avxco} increases by $\sim 15-30$ per cent for {\it L10}. At lower resolutions, this inclusion increases {\avxco} up to 50 per cent.
	In any case, it is reassuring that the results for {\it L10} are within the variations found for different nearby and resolved molecular clouds (see \citealt{Strong96}; \citealt{Dame01}; \citealt{Ripple13}; \citealt{Kong15} and many others), in particular because the SILCC-Zoom simulations are performed assuming solar neighbourhood conditions. We note that at later evolutionary times star formation would begin, forming massive stars, and stellar feedback would start to disperse the molecular clouds \citep{Haid19}. This particularly affects the cold molecular gas and hence {\avxco} would drop due to star formation and feedback \citep{Seifried20}.

	\begin{figure}
		\includegraphics[width=\columnwidth]{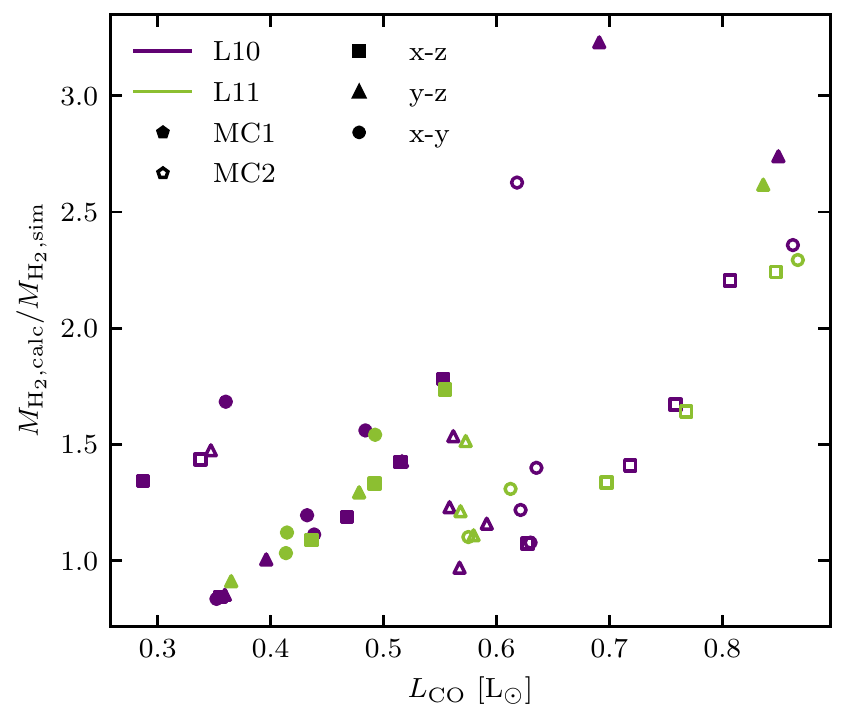}
		\caption{Plot of the {\molh} mass ratio, where $M_{\rm H_2,calc}$ is the {\molh} mass derived from the total, observable CO luminosity and \mbox{$ X_{\mathrm{CO,MW}}=2\times 10^{20}\:{\rm cm}^{-2}\:({\rm K\: km\: s}^{-1})^{-1}$ and $M_{\rm H_2,sim}$} is the total {\molh} mass obtained from the simulation data (see top row in Fig.~\ref{fig:M_co_h2}), against the observable CO luminosity $L_{\rm CO}$. The different colours show the resolutions {\it L10} and {\it L11}. Filled markers indicate MC1 and empty markers MC2. The different symbols indicate the respective LOS. Note that the CO luminosity increases from $t=1.5-2$ Myr and decreases from then on (as seen in Fig.~\ref{fig:X_CO_analysis_MC1} and \ref{fig:X_CO_analysis_MC2} in Appendix~\ref{app:xco}) such that more evolved clouds are on the left-hand side of the figure.}
		\label{fig:M_h2}
	\end{figure}
	
	From $X_{\mathrm{CO,MW}}$ and the total, observable CO luminosity, we calculate the {\molh} mass of the clouds for different times and LOS and compare them to the actual {\molh} mass present in the simulation. We compare the calculated $M_{\rm H_2,calc}$ to the total {\molh} mass (including CO-dark regions), as this is one of the arguments in favour of using $X_{\mathrm{CO,MW}}$ as discussed above. The results are presented in Fig.~\ref{fig:M_h2} for the best resolution runs {\it L10} (blue markers) and {\it L11} (orange markers). The different symbols indicate the LOS while the filled markers are for MC1 and the empty markers are for MC2. 
	The plot shows that the {\molh} mass is generally overestimated by a factor of 2. More diffuse (young) clouds have less CO-dark gas and hence the overestimation is larger \citep[up to a factor of $\sim$5; see also][]{Seifried20}. We note that the total CO luminosity initially increases and then decreases after $t=2$~Myr, as shown in Figs.~\ref{fig:X_CO_analysis_MC1} and \ref{fig:X_CO_analysis_MC2} in Appendix \ref{app:xco}. Hence, more evolved clouds are on the left side of the figure. The difference between {\it L10} and {\it L11} is small. Hence, independent of the numerical resolution, $X_{\mathrm{CO,MW}}$ is too high to be used for the solar neighbourhood clouds and simply overestimates the clouds' {\molh} mass.
	
	\begin{figure*}
		\includegraphics[width=\textwidth]{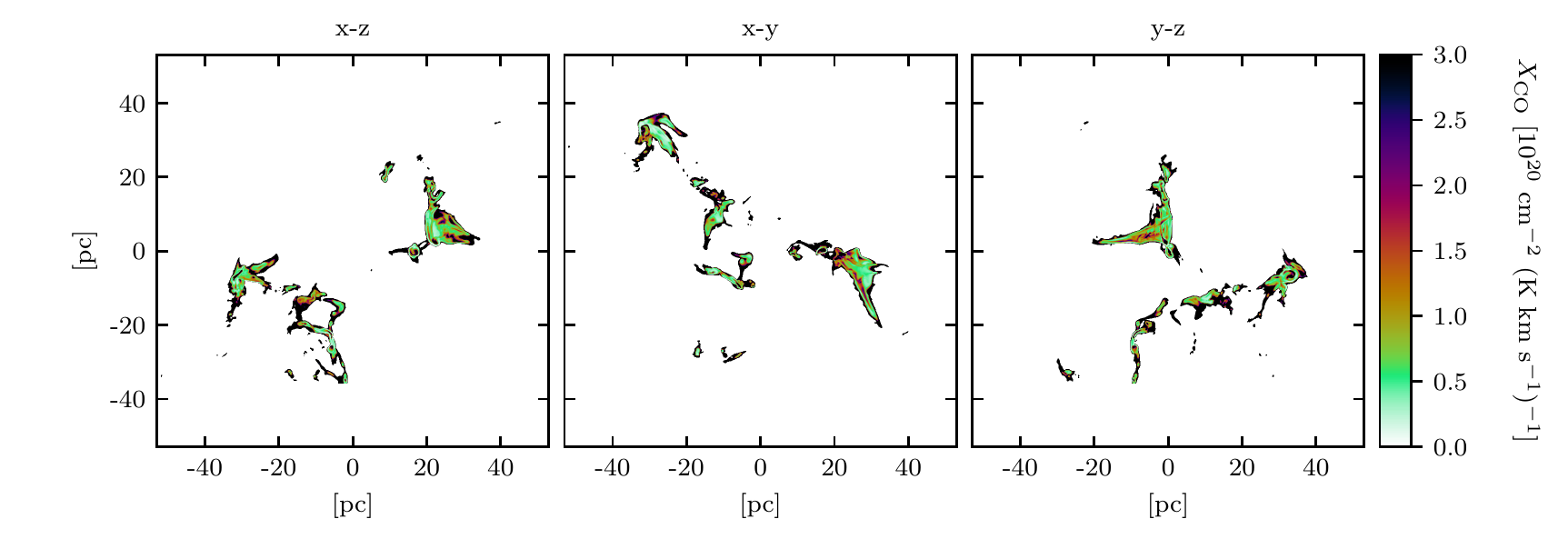}
		\caption{Local {\xco} factor of the observable regions of \textit{MC2\_L10} at 4 Myr. These maps show the spread of the {\xco} factor in the observable region $A_{\mathrm{obs,CO}}$. There is a considerable area with low {\xco}$\sim 0.5\times 10^{20}\:{\rm cm}^{-2}\:({\rm K\: km\: s}^{-1})^{-1}$. 
		While the majority of the map shows quite low results, the overall result for the whole regions is higher by a factor of 3-4, as the spread in of {\xco} is more than order of magnitude, as seen in the next figure and \protect\cite{Seifried17b}. Note that we show a linear scale which is cut off at $3\times 10^{20}\:{\rm cm}^{-2}\:({\rm K\: km\: s}^{-1})^{-1}$.}
		\label{fig:X_co_map}
	\end{figure*}
	
	\begin{figure*}
		\includegraphics[width=\textwidth]{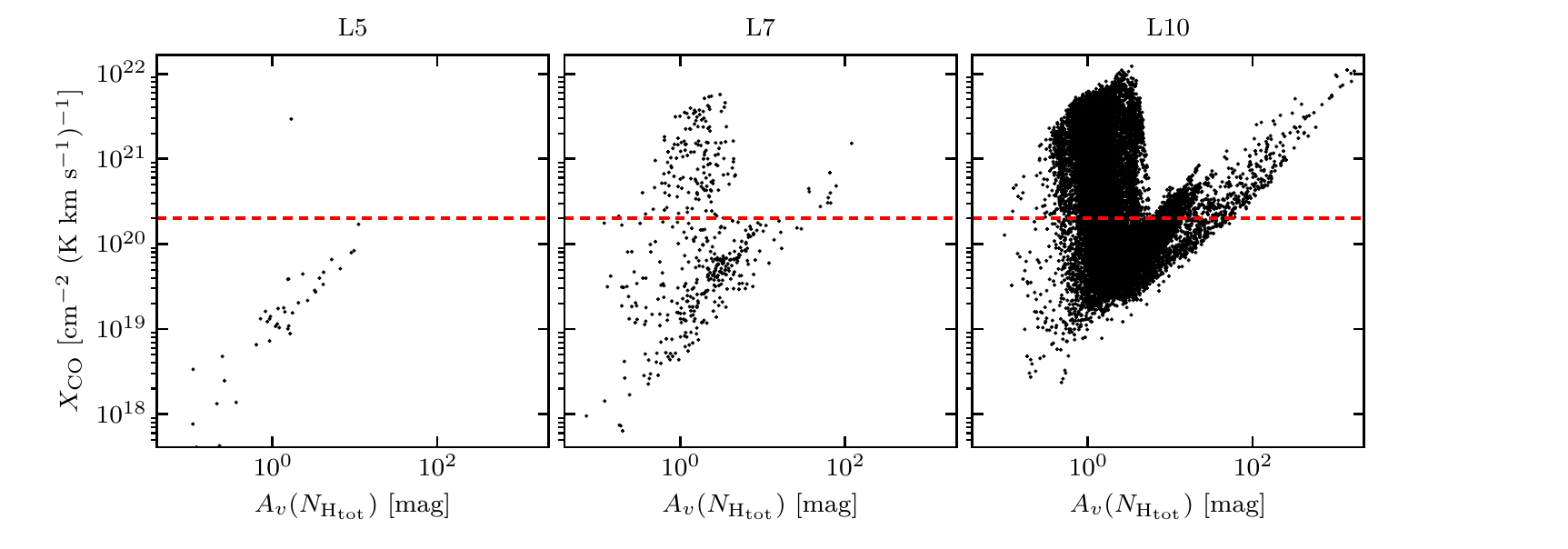}
		\caption{Distribution of the local {\xco} against $A_v(N_{\mathrm{H_{tot}}})$ for \textit{MC2\_L5} (left), \textit{MC2\_L7} (middle) and \textit{MC2\_L10} (right) at 4 Myr. We only show the results for the LOS along the y-axis as we find the same trend in each direction. The red dashed line indicates the location of \mbox{$ X_{\mathrm{CO,MW}}=2\times 10^{20}\:{\rm cm}^{-2}\:({\rm K\: km\: s}^{-1})^{-1}$}. Overall we see the generally observed trend of a decreasing local {\xco} factor for $A_v<3$ mag and an increasing trend for $A_v>3$ mag, though the latter is not quite present in the lowest resolution ({\it L5}) because the high column densities are unresolved. The decreasing {\xco} factor for $A_v<3$ mag is where the CO-dark molecular gas fraction is very high \citep[see][their Figure 8]{Seifried20}. The increasing {\xco} factor for $A_v>3$ mag results from CO quickly becoming optically thick.}
		\label{fig:Xco-Av}
	\end{figure*}

	\subsection{Pixel-based {\xco}-factor} \label{sec:resolved_xco}

    As {\avxco} approaches $X_{\mathrm{CO,MW}}$ at 4 Myr (see Fig.~\ref{fig:X_CO_analysis_MC2} in Appendix \ref{app:xco}), we take a closer look at the resolved {\xco} factor at this time for run {\it MC2\_L10}. 
    In Fig.~\ref{fig:X_co_map} we show maps of {\xco} for the three different LOS within the observable areas. The local {\xco} factor varies significantly across the map. These variations can be as large as a factor of 1000, as previously shown in \cite{Seifried17b}, their Figure 17 (note that we do not show a log scale, and that the colour bar is cut off at \mbox{$3\times 10^{20}\:{\rm cm}^{-2}\:({\rm K\: km\: s}^{-1})^{-1}$}). A first glance at the plots and a look at the column density in Fig.~\ref{fig:ratio-MC2} gives the impression that the factor is lowest in the central parts of the cloud, which are also the most dense, and highest in the diffuse regions near the edge of the observable area. This would be in apparent contrast with previously presented results, which suggest an increase of {\xco} with the local column density \citep[see e.g.][]{Sofue20}. 
    
    We therefore calculate the local visual extinction, $A_v$, from the total gas column density ($A_v = N_{\rm H,tot}/(1.87\times 10^{21}\:{\rm cm}^{-2}$); see \citealt{Walch15}) and create a scatter plot of the local {\xco} vs. $A_v$ in Fig.~\ref{fig:Xco-Av}. The figure shows three different spatial resolutions: {\it L5}, {\it L7} and {\it L10}. We present only the results for the LOS along the y-axis as all three LOS are comparable. We can reproduce the expected trend of the local {\xco} factor perfectly well: in regions with $A_v\lesssim 3$~mag there is a sharp decrease of {\xco} with $A_v$, while for $A_v\gtrsim 3$~mag {\xco} increases with increasing visual extinction. The decrease at low $A_v$ can be explained with significant fractions of CO-dark molecular gas \citep[see][their Figure 8]{Seifried20}. The increase at $A_v\gtrsim 3$~mag is caused by CO becoming optically thick in the dense regions. This reproduces recent observations \citep{Lee14,Sofue20} as well as previous numerical results \citep[see e.g.][]{Ossenkopf02, Bell06, Bell07, Clark15, Glover16, Szuecs16} and clears up the apparent contrast we found in Fig.~\ref{fig:X_co_map}. 
    
    From Fig.~\ref{fig:Xco-Av} it is also obvious why none of our derived quantities is converged for a spatial resolution of less than 1 pc: the high column density regions are not resolved at all and hence the calculated {\avxco} is dominated by regions which are not fully molecular. The small number of pixels at higher column densities could additionally lead to a high statistical noise.
	
	%%%%%%%%%%%%%%%%%%%% Xco FACTOR %%%%%%%%%%%%%%%%%%

	\section{Discussion of the {\xco} factor} \label{sec:Xco}
	The resolution study presented in Section~\ref{sec:emission_analysis} shows that for the synthetic observations of the CO emission of two molecular clouds formed from a supernova-driven, multi-phase ISM in the SILCC-Zoom simulations, convergence is only about to be achieved at the highest available spatial resolution {\it L11}, corresponding to d$x=0.06$~pc. From the differences seen between the two highest resolution levels ({\it L11-L10}), we estimate that convergence, if not yet achieved, could be achieved with another one or two refinement levels above {\it L11} (i.e. for d$x\approx0.01$~pc). However, the differences in all quantities are small for d$x\lesssim0.01$~pc.
	
	This is in accordance with the findings in \cite{Joshi19}, who showed for their simulations of molecular clouds forming in a colliding flow, that a spatial resolution of about 0.04 pc is required in order to reach convergence in the CO gas mass at any given time (as CO is forming together with the molecular cloud itself). We use Eq.~50 of \cite{Joshi19} to determine the required spatial resolution for convergence of CO in our simulation as
    \begin{equation}
    \mathrm{d}x_{\mathrm{CO}}\le 0.5~\mathrm{pc}\left(\frac{1}{\mu N_J}\right)^2\left(\frac{T_{\rm gas}}{10~\mathrm{K}}\right) \left(\frac{\left\langle\sigma \right\rangle_{\mathrm{mass}}}{1~\mathrm{km~s^{-1}}}\right)^{-1},
    \label{eq:convergence}
    \end{equation}
    with the mean molecular weight $\mu=2.35$ and the number of grid cells per Jeans length $N_J=1$\footnote{Note that typical simulations of star formation use at least 4 grid cells per Jeans length in order to resolve gravitational fragmentation \citep{Truelove97}, but here we rather use $N_J$ to estimate whether non-self-gravitating density peaks are resolved in the gas.}. The mass weighted velocity dispersion for the SILCC-Zoom simulation is obtained from \cite{Seifried18}, with values of \mbox{$\left\langle\sigma \right\rangle_{\mathrm{mass}}\in \left[3,~4.5\right]$ km s$^{-1}$}. For the gas temperature we choose a range expected for CO forming gas at \mbox{$T_{\rm gas}$ $ \in \left[10,~20\right]$ K}. 
    This results in the upper limit d$x_{\rm CO}=0.06$ pc for \mbox{$T_{\rm gas}$ $=20$ K} and $\left\langle\sigma \right\rangle_{\mathrm{mass}}=3$ km s$^{-1}$ and the lower limit of d$x_{\rm CO}=0.02$ pc with \mbox{$T_{\rm gas}$ $=10$ K} and $\left\langle\sigma \right\rangle_{\mathrm{mass}}=4.5$ km s$^{-1}$. The derived values are in agreement with our findings and with the expectation that {\it L11} or even one or two resolution levels more are necessary to obtain a converged result on the CO mass formed and, in turn, on $L_{\rm CO,obs}$ and {\avxco}.
    
    Our findings are hence not in agreement with the conclusion of \cite{Gong18} who state that a numerical resolution of 2 pc is required for {\xco} to be converged. The results presented here show that {\avxco} is slowly converging for d$x\lesssim 0.1$~pc (see Fig.~\ref{fig:deviations}).
    This difference may result from the different techniques to model the formation of CO. While \cite{Gong18} calculate the chemical abundance in a post-processing step by evolving the chemistry for 50 Myr, we model the CO formation on-the-fly in the SILCC-Zoom simulation, and the cloud formation time scale is much shorter than 50 Myr. In fact, 50 Myr even goes beyond the typical lifetime of a molecular cloud estimated to be rather of the order of $\sim$20 Myr (e.g. \citealt{Chevance19}). Due to this long time span of 50 Myr used, the abundances in \cite{Gong18} are likely to be in chemical equilibrium.  Hence, all dense and well shielded gas will be fully molecular and all resolutions will have reached the maximum CO fraction possible. 
    Furthermore, molecular clouds are non-static, dynamically evolving objects, e.g. due to turbulent mixing {\molh} might also not be in equilibrium \citep{Valdivia16, Seifried17b}. Overall, we argue that chemical post-processing over time spans significantly longer than typical cloud lifetime will result in significantly different synthetic observations than models which consider the dynamics and the chemistry of the molecular gas at the same time \citep[see e.g.][]{Glover10, Walch11, Walch15, Valdivia16, Seifried17b}.
   
    For the analysed times and the continuously available highest resolution SILCC-Zoom data {\it L10}, we obtain \mbox{$\left\langle X_{\mathrm{CO}}\right\rangle\approx 0.50-1.65~\times 10^{20}\:{\rm cm}^{-2}\:({\rm K\: km\: s}^{-1})^{-1}$} for MC1 and \mbox{$\left\langle X_{\mathrm{CO}}\right\rangle\approx 0.44-1.91~\times 10^{20}\:{\rm cm}^{-2}\:({\rm K\: km\: s}^{-1})^{-1}$} for MC2. In particular our young molecular clouds have $ X_{\mathrm{CO}}\lesssim 10^{20}\:{\rm cm}^{-2}\:({\rm K\: km\: s}^{-1})^{-1}$, so at least a factor of two smaller than the Milky Way average \citep{Bolatto13}. However, our values for {\avxco} are still within the variations of a factor of $\sim 2$ found for various different nearby clouds (see \citealt{Strong96}; \citealt{Dame01}; \citealt{Ripple13}; \citealt{Kong15} and many others), which is reassuring because the used simulations aim to reproduce molecular clouds forming in the solar neighbourhood. So, using \mbox{$ X_{\mathrm{CO,MW}}$}, the {\molh} mass is overestimated for the simulations presented here. However, \mbox{$ X_{\mathrm{CO,MW}}$} combines molecular clouds in different environments and different evolutionary stages. Star formation and feedback, magnetic fields, the gas metallicity, as well as other, higher gas densities, a higher cosmic ray ionisation rate or a higher impinging ISRF could all affect the resulting {\xco} factor, as e.g. discussed in \cite{Gong20}.
    
    In any case, it is essential to solve the CO chemistry and the dynamical evolution of the molecular cloud at the same time, and it is crucial to carry out the simulation at a high spatial resolution, such that the dense, molecule forming regions are well resolved. Otherwise the resulting synthetic observations are inaccurate as the radiative transfer calculations are based on unresolved simulation data.

	%%%%%%%%%%%%%%%%%%%% CONCLUSION %%%%%%%%%%%%%%%%%%

	\section{Conclusions} \label{sec:Conclusion}
	We present and analyse the synthetic CO and {\COthird} emission resulting from two molecular clouds which are forming from the supernova-driven, multi-phase interstellar medium as modelled within the SILCC and SILCC-Zoom simulations. The AMR simulations include an on-the-fly chemical network which follows the formation of {\molh} and CO. For each cloud, the same AMR simulation was repeated multiple times, each time with a higher maximum  spatial resolution (from 3.9~pc ({\it L5}) to 0.06~pc ({\it L11})). We follow the clouds' formation and evolution over 4~Myr. The simulations are post-processed with the radiative transfer code \textsc{radmc-3d} to obtain the synthetic emission maps. 
	
	To obtain the level populations of CO and {\COthird}, only ortho-{\molh} and para-{\molh} are generally used as collisional partners. Here, we show that H and He also have to be included as collisional partners, as they have a significant impact on the CO emission. Typically, He only causes a $\sim2-6$ per cent increase in the luminosity, but H leads to an average increase in CO luminosity of $\sim7-13$ per cent and potentially increases of $\sim26$ per cent. The prominent increase when including atomic hydrogen is a result of the fact that the outer parts of the molecular clouds do not only contain {\molh} but a mixture of H and {\molh}, leading to a more excited state of CO. The inclusion of additional collision partners also leads to an increase of the observable area $A_{\mathrm{obs,CO}}$, which we define to include all pixels with an integrated CO intensity of \mbox{$ \ge 0.1\:\mathrm{K\: km/s} $}. If observations allow for a higher sensitivity, considering He and H becomes even more important.
	
	We combine the synthetic emission maps of CO and {\COthird} to calculate the column density of CO, $N_{\rm CO}$. These calculations require estimates of the optical depth of the CO line $\tau_{\rm CO}$, which in turn requires an excitation temperature $T_{\rm ex}$. We investigate three different approaches which are commonly used to calculate the excitation temperature. We either set a fixed value for the whole region, calculate the excitation temperature from the peak intensity along the velocity space for each pixel, or take an intensity-weighted average of the pixel-wise excitation temperature. The latter is similar to using a fixed value but the value is tailored to the observations. By comparing the derived $T_{\rm ex}$ with the actual excitation temperatures calculated by \textsc{radmc-3d}, we find that the intensity-weighted average value is reasonable. We may then calculate $N_{\rm CO}$ and from it, the total CO mass of the cloud. We find that the intensity-weighted average excitation temperature actually gives CO masses which are in better agreement with the actual CO mass (derived from the simulation data) than the pixel-by-pixel values $T_{\rm ex,obs}$. The reason is that the pixel-by-pixel value leads to an overestimation of $N_{\rm CO}$ below $N_{\rm CO}\lesssim10^{18}\:{\rm cm^{-2}}$, while $N_{\rm CO}$ is underestimated at column densities above $N_{\rm CO}\gtrsim10^{19}\:{\rm cm^{-2}}$. The reason is that the calculation of $\tau_{\rm CO}$ is based on the assumption that CO is optically thick while {\COthird} is optically thin everywhere. This does not hold true for all pixels. It would be better to use an even more optically thin tracer than {\COthird} to derive the optical depth and hence the CO column density, or, if possible combine multiple different tracers to sample different column density regimes.
	
	When calculating the {\xco} factor from the synthetic emission and the {\molh} mass present in the simulations, we find that {\avxco} varies significantly between different spatial resolutions, clouds, lines of sight and simulation times. For our two clouds, the derived {\avxco} is somewhat lower than $X_{\mathrm{CO,MW}}$. More importantly, we investigate how the resolved {\xco} depends on the local gas column density across the maps. We recover the observed trend that {\xco} is decreasing with increasing $A_v$ for $A_v<3$ mag, while it is increasing with $A_v$ for $A_v>3$ mag (see e.g. \citealt{Lee14}; \citealt{Sofue20}). The reason is that only a fraction of the total carbon is in CO at low $A_v$, while CO quickly becomes optically thick at high $A_v$.
	
	Furthermore, we show that when including the full non-equilibrium chemistry in the simulation, the {\xco} factor is not converged at a resolution of 2 pc, as suggested by \cite{Gong18}. It is not even fully converged at a spatial resolution of d$x=0.06$ pc (our {\it L11} simulation), though from the trend seen in Fig.~\ref{fig:deviations} and a calculation done following \cite{Joshi19} it seems that convergence is close at this point. Compared to our highest resolution simulation, we find deviations in {\avxco} of $\sim$40 per cent for d$x=1.0$ pc, and $<10$ per cent for d$x\sim0.1$ pc. Even for d$x\sim0.1$ pc there are still in the {\molh} mass as well as in the CO luminosity. 
	
	\section*{Data availability}
    The data underlying this article will be shared on reasonable request to the corresponding author.
	
	\section*{Acknowledgements}
	We thank the referee, Prof. Liszt, for a very constructive report which greatly helped us to improve the paper.	EMAB and SW thank the Bonn-Cologne-Graduate School for their financial support. EMAB acknowledges the support of a Research Training Program scholarship from the Australian government as well as the Monash International Tuition Scholarship from Monash University. SW, SDC, and PN gratefully acknowledge the European Research Council under the European Community's Framework Programme FP8 via the ERC Starting Grant RADFEEDBACK (project number 679852). SW and AF further thank the Deutsche Forschungsgemeinschaft (DFG) for funding through SFB~956 ''The conditions and impact of star formation'' (sub-project C5). DS acknowledges the DFG for funding through SFB~956 ''The conditions and impact of star formation'' (sub-project C6). We would also like to acknowledge the people developing and maintaining the following open source packages which have been used extensively in this work: \textsc{Matplotlib} \citep{matplotlib}, \textsc{NumPy} \citep{numpy}, \textsc{SciPy} \citep{scipy}, \textsc{AstroPy} \citep{astropy} and \textsc{CMasher} \citep{cmasher}.
	%%%%%%%%%%%%%%%%%%%%%%%%%%%%%%%%%%%%%%%%%%%%%%%%%%
	
	%%%%%%%%%%%%%%%%%%%% REFERENCES %%%%%%%%%%%%%%%%%%
	
	% The best way to enter references is to use BibTeX:
	
	\bibliographystyle{mnras}
	\bibliography{bib}
	
	%%%%%%%%%%%%%%%%%%%%%%%%%%%%%%%%%%%%%%%%%%%%%%%%%%
	
	%%%%%%%%%%%%%%%%% APPENDICES %%%%%%%%%%%%%%%%%%%%%
	
	\appendix
	
	\section{Collision rates}\label{app:collision_rates}
	\begin{figure}
		\includegraphics[width=\columnwidth]{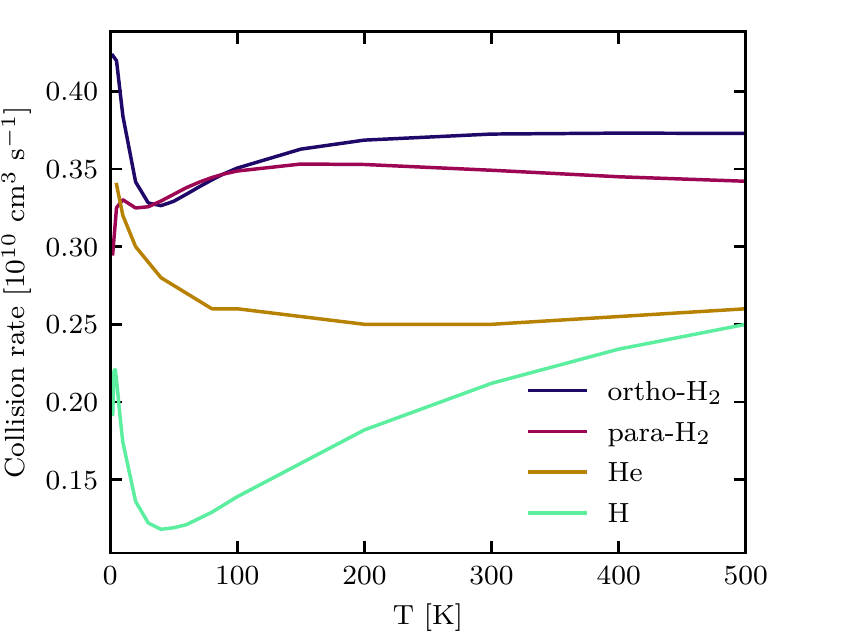}
		\caption{Collision rates at different temperatures for CO. Included are the rates of para-H$_2$ and ortho-H$_2$ from \protect\cite{Yang10}, He from \protect\cite{Cecchi02} and  H from \protect\cite{Walker15}.}
		\label{fig:collision_rates}
	\end{figure}
	In our simulations we use different collision rates for CO which we gathered from multiple sources. First we have the collision rates of the molecules with both para-{\molh} and ortho-{\molh} from \cite{Yang10} which we obtained from the LAMBDA database \citep{Schoier05}. We also obtained the collision rates for CO with He \citep{Cecchi02} and H \citep{Walker15}. Fig.~\ref{fig:collision_rates} shows the collision rates against the temperature.
	
	\section{Ratio of intensity maps with different collision partners} \label{app:collision}
	\begin{figure*}
		\includegraphics[width=\textwidth]{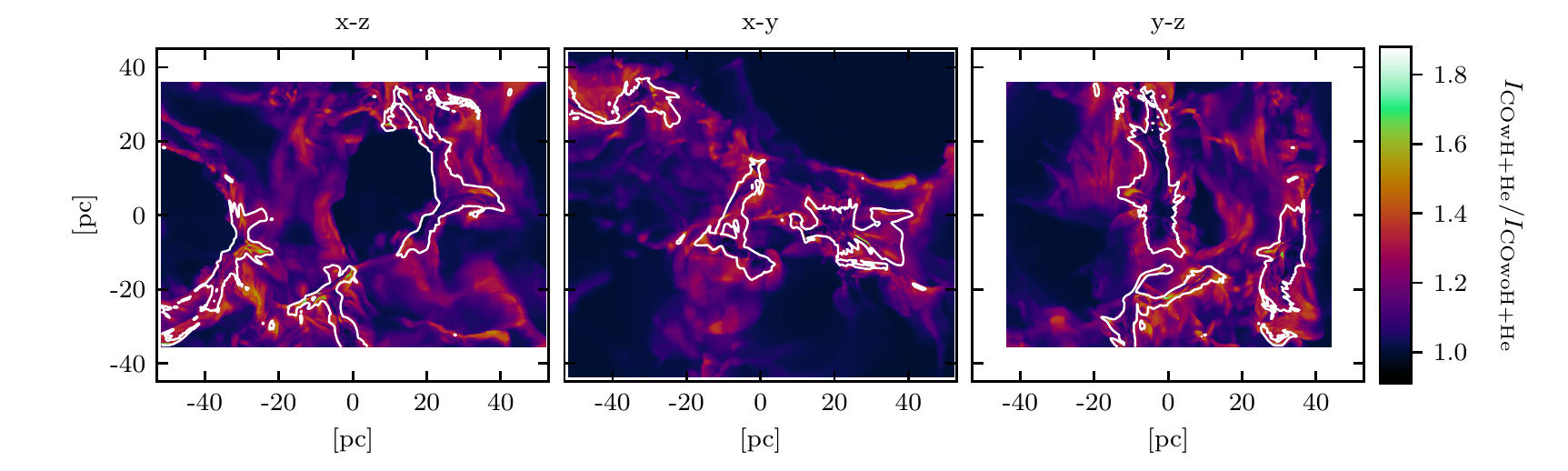}
		\caption{Ratio of the integrated intensity maps including He and H as collision partners ($I_{\rm COwH+He}$) and the maps excluding them ($I_{\rm COwoH+He}$). Shown are the ratios for {\it MC2\_L10} at 2 Myr. The white contour shows the area $A_{\mathrm{obs,CO}}$ for the intensity maps including the additional collision partners. The changes in the dense regions of the cloud stem from the inclusion of He, while the changes in the diffuse regions are from the inclusion of H as hydrogen has not yet gone fully molecular in these regions.}
		\label{fig:ratio_collision}
	\end{figure*}
	
	\begin{figure*}
		\includegraphics[width=\textwidth]{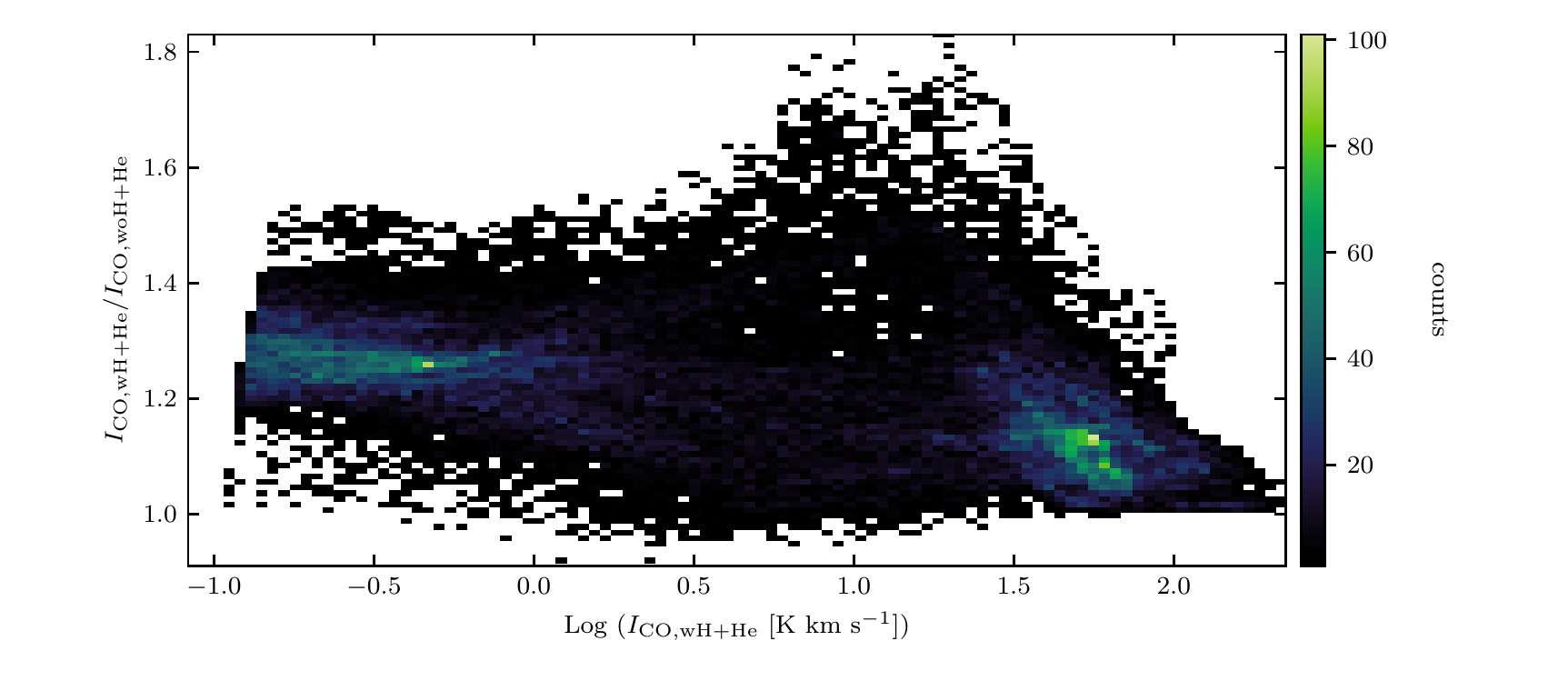}
		\caption{2D PDF of the ratio of the integrated intensity maps including He and H as collision partners ($I_{\rm COwH+He}$) and the maps excluding them ($I_{\rm COwoH+He}$) against the integrated intensity with additional collision partners ($I_{\rm COwH+He}$), showing that the effect of including additional collision partners affects the most diffuse and dense regions.}
		\label{fig:ratio_hist}
	\end{figure*}
	
	Fig.~\ref{fig:ratio_collision} shows the ratio of the integrated intensity maps including additionally the collision partners of He and H and excluding them for the run \textit{MC2\_L10} at 2 Myr for all three lines of sight. White contours on the maps indicate where the integrated intensity of the map including the additional collision partners is $ \ge 0.1\:\mathrm{K\: km\: s^{-1}} $. The bulk of the changes that occur outside as well as towards the edges inside of the white contours are a result of the inclusion of H as a collision partner. H therefore seems to be an important collision partner especially in the diffuse parts of the molecular clouds as these are regions where not all hydrogen has yet become molecular. The changes that can be observed especially within the dense regions are a result of the inclusion of He as a collision partner.
	
	To see more in depth where the additional collision partners affect the emission, we look at a 2D PDF of the ratio of the integrated intensity maps including He and H and the maps excluding them against the integrated intensity including additional collision partners in Fig.~\ref{fig:ratio_hist}. We can see that at the diffuse edge of the cloud the intensity is enhanced mostly by a factor of $\sim 50$ per cent. When the intensity is around \mbox{$10\:{\rm K\:km\:s^{-1}}$}, the inclusion of He and H change very little. For higher intensities around \mbox{$40\:{\rm K\: km\: s^{-1}}$} the intensity increases again, this time by a factor of $\sim 20$ per cent. Overall, including the additional collision partners yields to the increase of the observed luminosities and therefore the emission of the cloud. This happens when looking at the emission for the same area as seen in Fig.~\ref{fig:lum_col_partners}. As the additional collision partners also raise the emission in the diffuse parts of the clouds, the area for our observational threshold also increases, which results in a further increase of the luminosity as the observed area increases.
	
	\section{Time evolution of CO mass ratio for different $T_{\mathrm{ex}}$}\label{app:Tex}
	\begin{figure}
		\includegraphics[width=\columnwidth]{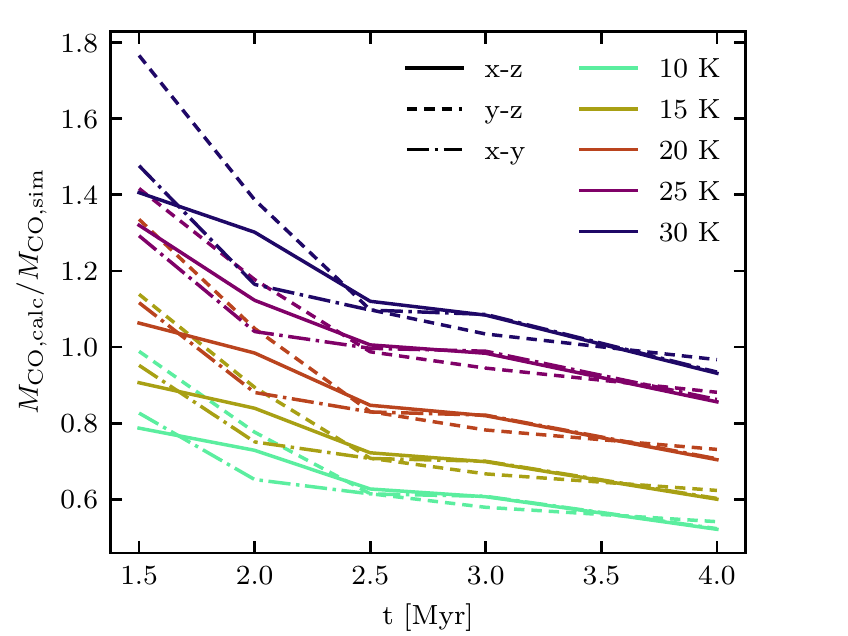}
		\caption{Time evolution of the total CO mass ratio from the calculated and the simulation column density. The ratios are for different $T_{\mathrm{ex,fix}}$. Shown are the results for \textit{MC2\_L10}. The individual excitation temperature trends show a decrease in the ratio over time for a fixed excitation temperature. The various $T_{\mathrm{ex,fix}}$ used here show that the ratio increases for increasing $T_{\mathrm{ex,fix}}$.}
		\label{fig:Nratio-t-MC2}
	\end{figure}
 	Fig.~\ref{fig:Nratio-t-MC2} shows how the calculation of the CO column density behaves compared to the present CO column density when using different fixed excitation temperatures in the calculations. We choose to look at the excitation temperature in the range of 10 - 30 K, as these are values that can typically be found for the excitation temperature in molecular clouds where star formation has not yet started. All calculations show a similar trend for the same LOS, with the excitation temperature resulting in a shift of the results along the y-axis. Overall there is also a decreasing trend with time, which indicates that higher excitation temperatures are required as more time passes. 
	
	\section{Relative changes between resolution levels} \label{app:differences}
	\begin{figure*}
		\includegraphics[width=\textwidth]{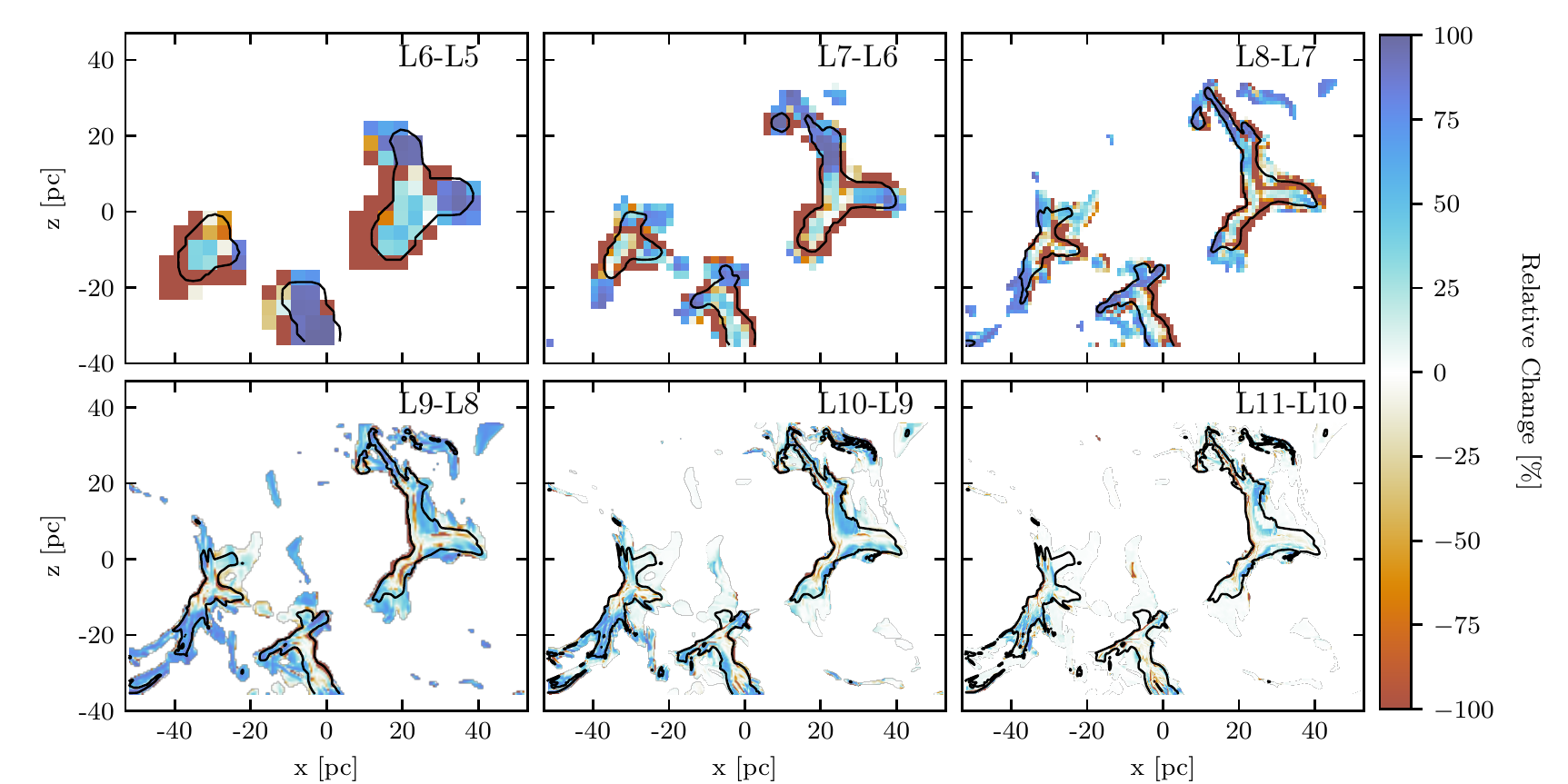}
		\caption{Relative changes of the integrated intensity maps from the CO(1-0) transition for different spatial resolutions, as indicated in the top right corner of each sub-map. The changes are given relative to the higher resolution level. All maps show the differences for the front view of MC2 at 2 Myr. The black contour depicts the observable region $A_{\mathrm{obs,CO}}$ at the higher resolution level. }
		\label{fig:lvl-diff-MC2}
	\end{figure*}
	In order to analyse the changes of the integrated intensity as a function of the spatial resolution of the SILCC-Zoom simulation data, we look at the relative changes between the resolution levels given in percentage of the higher level in Fig.~\ref{fig:lvl-diff-MC2}. The sub-maps show the differences of different levels for the front view of MC2 at 2 Myr. For the lower resolution levels, the differences between the levels are larger than 100 per cent, while for higher levels the differences stay within the confines of the cloud. At the highest two levels, regions of the map with very low integrated intensities show that the lower level has the higher emission. This is not reproduced in the previous levels and indicates that for the regions of low density and low emission, convergence has been achieved, though these lie outside of the areas we assume to be observable $A_{\mathrm{obs,CO}}$. Within $A_{\mathrm{obs,CO}}$, optically thin regions still show a more intense emission at higher resolution. Presumably this reflects the finer density substructure of the turbulent cloud in the better resolved case.
	
	\section{Analysis of the changes in {\avxco}} \label{app:xco}
	\begin{figure*}
		\includegraphics[width=1\textwidth]{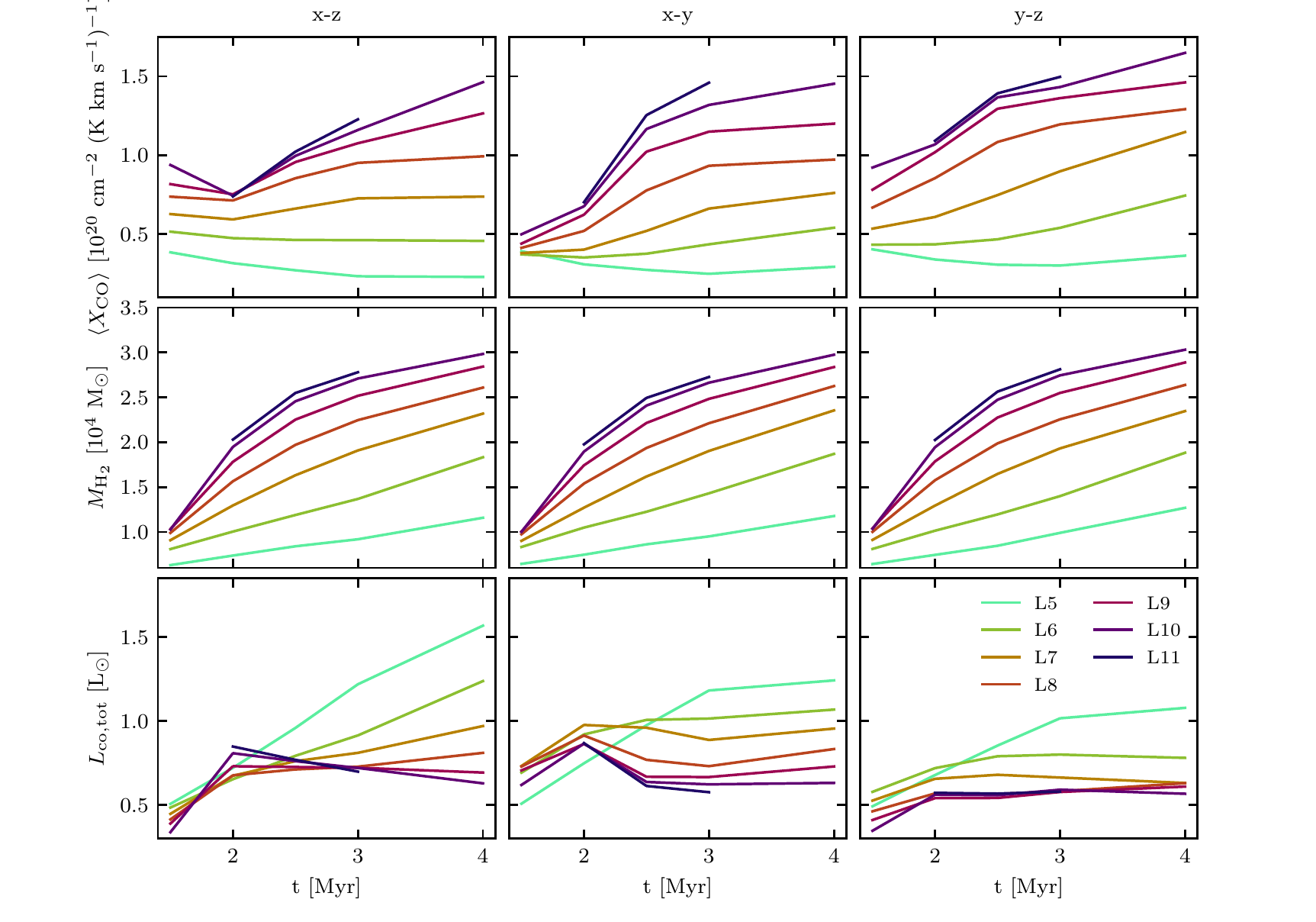}
		\caption{Analysis of the change in $\left\langle X_{\mathrm{CO}}\right\rangle$ in the observable area $A_{\mathrm{obs,CO}}$ for all three LOS of MC1. Shown are three different time evolutions: Top $\left\langle X_{\mathrm{CO}}\right\rangle$, middle the {\molh} mass, bottom the total luminosity of CO form the observable regions. The different colours indicate the different resolutions. From left to right are the LOS along the y-axis, z-axis and x-axis.}
		\label{fig:X_CO_analysis_MC1}
	\end{figure*}
	\begin{figure*}
		\includegraphics[width=1\textwidth]{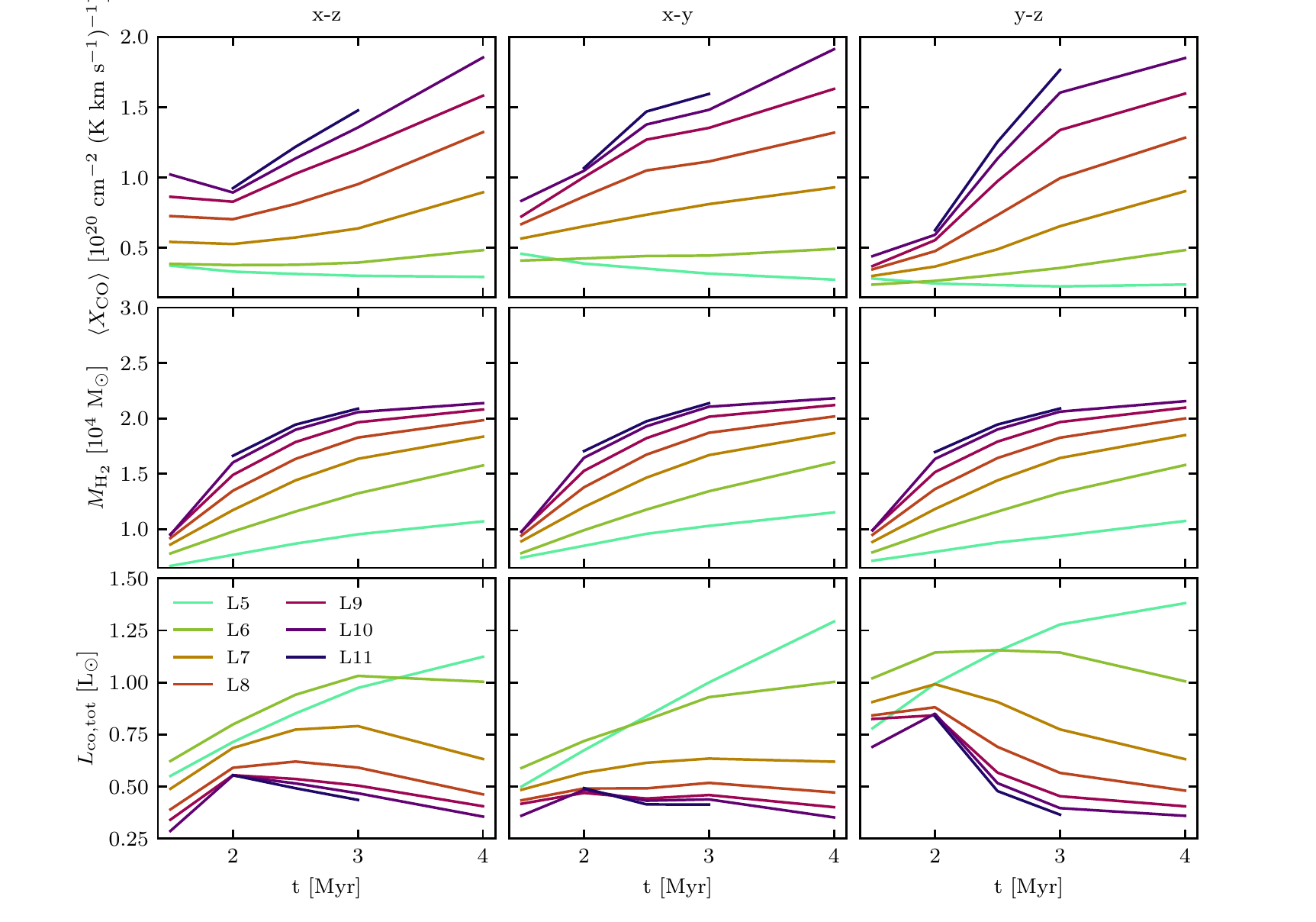}
		\caption{Same as Fig.~\ref{fig:X_CO_analysis_MC1} but for MC2}
		\label{fig:X_CO_analysis_MC2}
	\end{figure*}
	Figures \ref{fig:X_CO_analysis_MC1} and \ref{fig:X_CO_analysis_MC2} show the time evolution of {\avxco} as well as its contributing factors $M_{\mathrm{H_2}}$ and $L_{\mathrm{CO}}$ for each view and resolution level in the observable regions of $A_{\mathrm{obs,CO}}$. The figures show that the changes of $M_{\mathrm{H_2}}$ and $L_{\mathrm{CO}}$ reflect in the results of {\avxco}.
	
		\section{{\avxco} of the whole map}\label{app:x_co_of_map}
	\begin{figure*}
		\includegraphics[width=\textwidth]{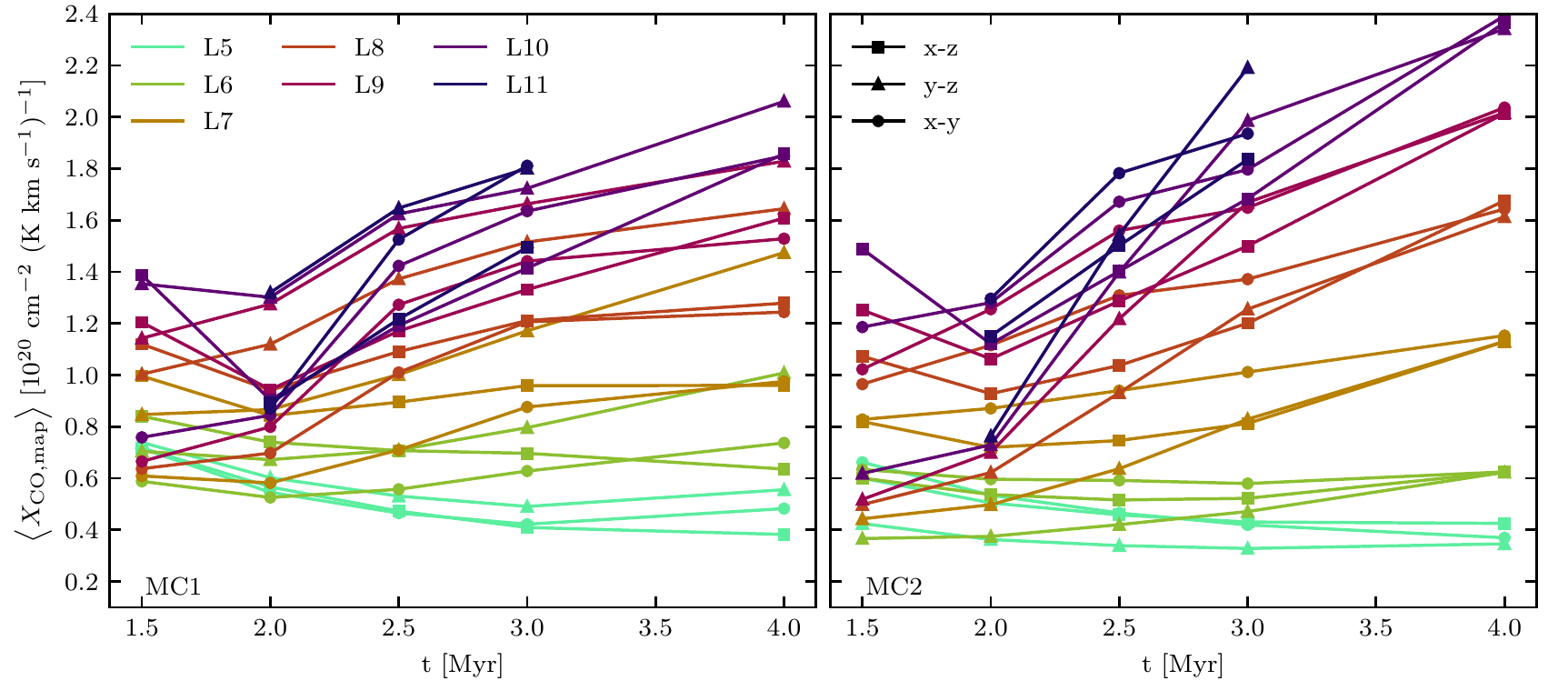}
		\caption{{\avxco} factor of the whole map for MC1 (left) and MC2 (right) over time. The different colours represent the different spatial resolutions as introduced in Table \ref{tab:refinement}, while the symbols depict the different projections. Overall, the factor increases with increasing spatial resolution. The values are higher for the whole map compared to only the observable regions shown in Fig.~\ref{fig:X_co}. This is due to the CO-dark gas present, which is not accounted for in the calculations of the observable region. For the whole map we have about 20 per cent more hydrogen while the luminosity increases only by about 2 per cent, leading to the differences in the results of {\avxco}.}
		\label{fig:X_co_of_map}
	\end{figure*}
    Contrary to real observations we also know how much {\molh} mass resides within the "CO-dark region" of the map. Thus we recalculate {\avxco} for the whole map. The results are shown in Fig.~\ref{fig:X_co_of_map}. This {\avxco} factor seems to be in reasonable agreement with observed results, which is mostly caused by the changes in the {\molh} mass as we now also include the CO-dark {\molh} in the calculations. About $15-30$ per cent of the {\molh} mass is CO-dark \citep[see also][]{Seifried20}, resulting in a $15-30$ per cent increase of the {\xco} factor when looking at the whole map instead of the observable region only; $25-30$ per cent at 1.5 Myr and then $15-20$ per cent for the remaining time steps. The total CO luminosity only changes by about 2 per cent when considering the whole map instead of the observable region.
	
	%%%%%%%%%%%%%%%%%%%%%%%%%%%%%%%%%%%%%%

	% Don't change these lines
	\bsp	% typesetting comment
	\label{lastpage}
\end{document}